\documentclass[11pt,a4paper]{article}
\pdfoutput=1
\usepackage{jheppub}
\usepackage{xcolor}

\newcommand{\sgn}{\mathrm{sgn}}
\newcommand{\be}{\begin{equation}}
\newcommand{\ee}{\end{equation}}
\newcommand{\bal}{\begin{aligned}}
\newcommand{\eal}{\end{aligned}}
\newcommand{\Log}{\mathrm{ln}}

\title{A Ground State for the Causal Diamond in 2 Dimensions}

\author[a,b]{\small Niayesh Afshordi,}
\author[c]{Michel Buck,}
\author[b,c,d]{Fay Dowker,}
\author[e]{David Rideout,}
\author[b,f]{ Rafael D. Sorkin,}
\author[a]{and Yasaman K. Yazdi}

\affiliation[a]{Department of Physics and Astronomy, University of Waterloo, Waterloo ON, N2L 3G1, Canada}
\affiliation[b]{Perimeter Institute for Theoretical Physics, 31 Caroline St. N., Waterloo ON, N2L 2Y5, Canada}
\affiliation[c]{Blackett Laboratory, Imperial College, London, SW7 2AZ, U.K.}
\affiliation[d]{Institute for Quantum Computing, University of Waterloo, Waterloo ON, N2L 3G1, Canada}
\affiliation[e]{Department of Mathematics, University of California San Diego, La Jolla CA, 92093-0112, U.S.A.}
\affiliation[f]{Department of Physics, Syracuse University, Syracuse NY, 13244-1130, U.S.A.}

\emailAdd{nafshordi@perimeterinstitute.ca}
\emailAdd{m.buck11@imperial.ac.uk}
\emailAdd{f.dowker@imperial.ac.uk}
\emailAdd{drideout@math.ucsd.edu}
\emailAdd{rsorkin@perimeterinstitute.ca}
\emailAdd{yyazdi@perimeterinstitute.ca}


\abstract{
We
apply
a recent proposal for a distinguished 
ground state 
of
a quantum field in a globally hyperbolic spacetime
to the
free massless scalar field in a causal diamond in two-dimensional
Minkowski space.
We investigate the two limits in which the Wightman
function is evaluated (i) for pairs of points that lie in the centre of
the diamond (i.e. far from the boundaries), and (ii) for pairs of points
that are close to the left or right corner.
We find that in the centre, the Minkowski vacuum state is recovered,
with a
definite
value of the
infrared cutoff.
Interestingly, the ground state is not the
Rindler vacuum
in
the corner of the diamond, as might have been
expected, but is instead the vacuum of a flat space in the presence of a
static mirror on that corner.
We confirm these results
by numerically evaluating
the Wightman function
of
a massless scalar field on a causal set corresponding to
the
continuum
causal diamond.
}

\begin{document}
\maketitle
\thispagestyle{empty}

\section{Introduction}

At the end of his classic textbook `Aspects of Quantum Field Theory in Curved Space-Time,' \cite{Fulling:1989nb}
S.A. Fulling points out the central tension in the subject as traditionally studied:

\begin{quote}
One of the striking things about the subject is the intertwining of the
conceptual issues with the mathematical tools. Historically there has
been a close association between relativity and differential geometry,
while rigorous research in quantum theory has looked more toward
functional analysis. Field quantisation in a gravitational background
brings these two alliances into head-to-head confrontation: A field is
a function of a time and a space variable,
$$\phi(t,\mathbf{x}).$$
Relativity and modern geometry persuade one with an almost religious intensity
 that these variables must be merged and submerged;
 the true domain of the field is a space-time manifold:
$$\phi(\underline{x}),\;\;\underline{x}\rightarrow\left\{x^\mu\right\}$$
But quantum theory and its ally, analysis, are constantly pushing in the opposite
direction. They want to think of the field as an element of some function space, depending
on time as a parameter:
$$\phi_t(\mathbf{x}).$$
\end{quote}

Taking the relativistic view that the physical world has a
space-time character indeed requires an approach to quantum field theory
that is built on spacetime concepts, one that makes no fundamental appeal
 to ``time as a parameter.''
The approach of algebraic quantum field theory takes this view seriously
and attempts to push the foundations
towards greater covariance,  basing the theory on
appropriate algebras of operators associated
to open spacetime regions. The focus on the quasi-local algebras as the
central constructs of the theory has encouraged the point of view that in general in
curved spacetime there is no preferred quantum state and that a choice of state is akin to
a choice
of coordinates (see {\textit{e.g.}} \cite{Wald:2009uh}).
However, the
algebraic approach has not
been
completely successful
in constructing the expectation value of the
stress-energy tensor or dealing with interacting fields.
It remains an open question whether
quantum field theory requires in addition the identification of a
distinguished ``ground state''
(or class of them).
Recently, a proposal has been made for exactly such a ``ground state of
a spacetime region'' for a free quantum field theory
\cite{Sorkin:2011pn, Siavash}.
The proposal, implicit in work by
S. Johnston on quantum field theory on a causal set
\cite{Johnston:2009fr}, gives primacy to the `true domain,' the
spacetime manifold itself and its coordinate independent causal
structure.
The Wightman or two-point function, $W(X,Y)$, of
this
``Sorkin-Johnston'' (SJ) ground state is defined in an essentially covariant way
starting
from
nothing more than
$\Delta(X,Y)$, the Pauli-Jordan function or
commutator function.
Since the SJ ground state is
taken to be
Gaussian,
its $n$-point functions are
then
completely determined
in terms of
$W$.

In this paper, we study the SJ proposal in the case of a free massless
scalar field in a causal diamond of two-dimensional Minkowski
spacetime. We pay particular attention to the Wightman function obtained
in the limit where the size of the diamond is large compared to the
geodesic separation between the two arguments of $W$, and where the
points lie either (i) far away from the 
left and right corners or (ii) close
to one of the corners.

In the former
limit
one might hope to recover the unique Poincar\'e-invariant
Minkowski vacuum state, if such a state existed.  But in fact, no such
vacuum exists, since $W$ is Logarithmic and depends on an
arbitrary length-parameter or ``infrared cutoff'', as is well known.
In our finite diamond, we find that $W$ has
the expected form, but with a definite value
of the length-parameter determined by the diamond's area.
%

In
the latter limit, one might expect to see the Rindler vacuum state, since
the geometry of the corner approaches that of the familiar Rindler wedge
as the diamond size tends to infinity. However, we find that this is not
the case. Instead, the SJ state close to the corner is the vacuum state
of Minkowski spacetime with a static mirror on the corner.

Further, we
use the original causal set QFT formulation \cite{Johnston:2009fr} to
construct the ground state on a causal set that approximates the
continuum
causal diamond. We compare the results with the foregoing continuum
analysis in the subregions of
the causal set
corresponding to the two
limits (i) and (ii).  In both cases, the Wightman function on the causal
set is in good agreement with the continuum Wightman function.\\

\section{The SJ vacuum}\label{sec_SJ}

We state the SJ proposal for the ``ground state'' of a free scalar field
in a $d$-dimensional globally hyperbolic
spacetime 
$(\mathcal M, g_{\mu\nu})$~\cite{Sorkin:2011pn, Johnston:2009fr}.
The starting point is the Pauli-Jordan function $\Delta  (X,X')$, where we use capital letters
$X, X'$ to denote spacetime points. It is defined by
\be
\Delta(X,X'):=G_R(X,X')-G_R(X',X)
\label{eq:pjdef}
\ee
where $G_R$ is the retarded propagator, i.e. the Green function of the
Klein-Gordon operator\footnote{The signature of the spacetime metric is
$(- + + \dots+)$ and $\Box = g^{\mu\nu} \nabla_\mu\nabla_\nu$.}
$\Box-m^2$
satisfying the retarded boundary condition,
$G_R(X,X')=0$ unless $X'\prec X$, meaning that $X'$ is to the causal
past of $X$.
Notice that $\Delta$ is real and antisymmetric.

The rules of
canonical quantisation relate
the Pauli-Jordan function
to the commutator of field operators by:
\be
 i \Delta  (X,X')
 = [ \hat\phi(X), \hat\phi(X')  ]   \ ,
  \label{eq:paulijordancommutator}
\ee
%
while
the Wightman two-point function of a state $|0\rangle$ is
\be
   W_0(X, X') = \langle 0| \hat\phi(X)\hat\phi(X')|0\rangle \,.
\ee
The Wightman function of the SJ state is defined by
the three conditions~\cite{Sorkin:2011pn}:
\begin{enumerate}
\item \emph{commutator: }  $i\Delta(X,X') = W(X,X')-W^*(X,X')$
\item \emph{positivity: } $\int_{\cal{M}} dV \int_{\cal{M}} dV' f^*(X) W(X, X') f(X') \ge 0$
\item \emph{orthogonal supports: }
  $ \int_{\cal{M}} dV'  \; W(X, X') \, W(X',X'')^*  = 0 $,
\end{enumerate}
where $\int dV = \int d^d X \sqrt{-g(X)}$.
On this view the Wightman function is a positive bilinear form on the
vector space of complex functions.
The first two conditions must be satisfied by the two-point function of any state. It is the third and
last requirement that acts as the ``ground state condition.''

For the conditions to specify a state fully,
their solution  must be unique.
Suppose $W_1$ and $W_2$ both satisfy the
conditions. Let us express the ground state condition
as $WW^* =0$\footnote%
{Strictly speaking, to multiply two quadratic forms together requires a metric,
  here given by a delta function. In these formulas the star
  denotes numerical complex-conjugate, not hermitian adjoint.}.
We have
$W_1 W_1^* = W_1^* W_1 = W_2 W_2^*= W_2^* W_2 = 0$ and $W_1 - W_1^* = W_2 - W_2^*$.
This implies
\begin{align*}
(W_1-W_1^*)^2 &= (W_2 - W_2^*)^2\\
\Rightarrow (W_1 + W_1^*)^2 & = (W_2 + W_2^*)^2\,.
\end{align*}
Both $W_1 + W_1^*$ and $W_2 + W_2^*$ are
positive bilinear forms,
and if there are unique positive
square roots of such forms then this implies
\begin{align*}
W_1+ W_1^* &= W_2 + W_2^* \\
\Rightarrow \ \ \ \ W_1 &= W_2\,.
\end{align*}

Formally, we can interpret the ground state condition as the requirement
that $W$ be the positive part of $i\Delta$,
thought of as an operator on the Hilbert space of square integrable
functions ${L}^{2}( \mathcal M,dV)$ \cite{Sorkin:2011pn}.
This allows us to describe a more or less direct construction of $W$
from the Pauli-Jordan function~\cite{Johnston:2009fr,Johnston:2010su,Siavash}.
The distribution $i\Delta(X,X')$ defines the kernel of a (Hermitian) integral operator
(which we may call the \emph{Pauli-Jordan operator} $i\Delta$)
on ${L}^{2}( \mathcal M,dV)$.  The inner product on this space is
\be
\langle f, g \rangle:=\int_{\mathcal M} dV f(X)^*g(X),
\ee
where $dV=\sqrt{-g(X)}d^d X$ is the invariant volume-element
on $\mathcal M$.
The action of $i\Delta:f\mapsto i\Delta f $ is then given by
\be
(i\Delta f)(X) := \int_{\mathcal M }^{ } dV' i\Delta(X,X') f(X').
\label{eq:defpjfunctional}
\ee
When this operator is self-adjoint, it admits a unique spectral
decomposition.
Whether or not this (or some suitable generalization of it) is the case
depends on the functional form of the kernel $i\Delta(X,X')$ and on the
geometry of $\mathcal M$.
For the massless scalar field on a bounded region of 
Minkowski space, such as the finite causal diamond which is the subject
of this paper, $i\Delta$ is indeed a self-adjoint operator, since the
kernel $i\Delta(X,X')$ is Hermitian and bounded (see~\eqref{36}
below). In the following we assume that
we can expand
$i\Delta$
in terms of its eigenfunctions.

Noting that the kernel $i\Delta$ is skew-symmetric, we find that the
eigenfunctions in the image of $i\Delta$ come in complex conjugate pairs
$T^+_q $ and $T^-_q $ with real eigenvalues $\pm \lambda_q $:
\be
(i\Delta T^\pm_q )(X)=\pm\lambda_q  T^\pm_q (X),
\ee
where $T^-_q (X)=[T^+_q (X)]^*$ and $\lambda_q >0$.
Now by the definition of $i\Delta$, these eigenfunctions must be
solutions to the homogeneous Klein-Gordon equation.
If they are $L^2$-normalised
so that $||T^+_q ||^2:=\langle T^+_q ,T^+_q \rangle=1$,\footnote%
{~When the spacetime region is non-compact this should be replaced by a
  delta-function normalisation $\langle T^+_q ,T^+_{q'}\rangle=\delta(q -q ')$.}
then the spectral decomposition of
the Pauli-Jordan function implies that the kernel can be written as
\be
i \Delta  (X,X' )= \sum_{   q }^{ }\lambda_{q } T^+_{q }(X)T^+_{q }(X')^*-\sum_q \lambda_{q } T^-_q (X)T^-_q (X')^\star.
\label{eq:SJpaulijordan}
\ee
We now construct the SJ two-point function ${W}_{SJ}(X,X')$ by restricting~\eqref{eq:SJpaulijordan} to its positive part:
\be
{W}_{SJ}(X,X'):=  \sum_{  q }^{ }\lambda _{q }T^+_{q }(X)T^+_{q }(X')^\star=\sum_{  q }^{ }\mathcal T_{q }(X)\mathcal T_{q }(X')^\star
\label{eq:SJwightman}
\ee
where ${\mathcal{T}}_q (X):= T^+_q(X)\sqrt{\lambda_q}$.

Let us compare this with the familiar case of the vacuum state
of a free scalar field when the spacetime admits a
timelike Killing vector $\kappa = \partial_t$.  The free field can be expanded in
a basis of  modes,  $\{u_k, u_k^*\}$,  which are positive and negative frequency
with respect to $\kappa$:
\be
 u_k(X) = u_k(t, \mathbf{X})
  =N_k e^{-i \omega_k t}
 U_k(\mathbf{X})\, ,
\ee
where $N_k$ is a normalisation-factor,
$\omega_k>0$,
 and $\mathbf{X}$ are the spatial coordinates of $X$.
When the emission and absorption operators that serve as the expansion
coefficients are given their customary normalisation,
then the Pauli-Jordan function can be expressed as a mode sum
\be
  i\Delta(X,X')=\sum_k\left[u_k(X)u^*_k(X')-u_k^*(X)u_k(X')\right]\,.
  \label{paulijordanmodeexpansion}
\ee
The vacuum state corresponding to the choice of positive frequency modes
$u_k$ is the ground state of the Hamiltonian associated to $\kappa$,
and its Wightman function is
\be
  W(X,X')=\sum_k u_k(X)u^*_k(X')\,.
\label{wightmanmodeexpansion}
\ee
Comparing~\eqref{wightmanmodeexpansion}
with  \eqref{eq:SJwightman}
we see the eigenfunctions $\mathcal T_q :=\sqrt{\lambda_q }T^+_q $ chosen
by the SJ condition are the analogues of the positive frequency modes in the
static spacetime,
but the SJ
construction
does not require the
existence of a timelike isometry.

When there {\textit{is}} a timelike Killing vector,
we can show that the SJ state
(formally extended to the case of a spacetime of infinite volume)
is the
ground state of the Hamiltonian associated with that Killing vector by showing that the
Wightman function defined by \eqref{wightmanmodeexpansion}  satisfies the
SJ conditions.
Clearly,
\eqref{wightmanmodeexpansion} satisfies conditions (i) and (ii),
and it only remains to check the ground state condition (iii). We have
\begin{align*}
 WW^* (X, X'') =& \int_{\cal{M}} dV'
 W(X, X') W^*(X', X'')  \\
=& \sum_k \sum_{l} u_k(X) u_{l}(X'')
\int dt  d^{n-1}\mathbf{X}' \sqrt{-g(X')} u^*_k(X') u^*_l(X')\\
= &
\sum_k \sum_{l} u_k(X) u_{l}(X'')
\int d^{n-1}\mathbf{X}' \sqrt{-g({\mathbf{X}}')}  U^*_k({\mathbf{X'}})U^*_l({\mathbf{X'}})
\int_{-\infty}^\infty dt e^{-i\omega_k t} e^{-i \omega_l t}\\
= &
\sum_k \sum_{l} u_k(X) u_{l}(X'')
\int d^{n-1}\mathbf{X}' \sqrt{-g({\mathbf{X}}')}  U^*_k({\mathbf{X'}})U^*_l({\mathbf{X'}})
\int_{-\infty}^\infty dt e^{-i (\omega_k + \omega_l) t}\\
\propto & \,\,\delta( \omega_k + \omega_l) = 0
\end{align*}
since the sum over modes does not include the zero-mode $\omega_k =0$.
Thus, if the SJ conditions specify a unique state, that state is the appropriate
ground state when there is a globally timelike Killing vector.

This might be a good place to comment on the question of whether the SJ
vacuum obeys the so-called Hadamard condition on its short-distance
behavior.  In static spacetimes the Hamiltonian vacuum obeys this
condition, so the SJ vacuum does as well, as we have just seen.  On the
other hand, Fewster and Verch \cite{Fewster} have recently provided
examples of regions where the condition does not hold.  In our opinion,
the significance of the Hadamard condition will not be known until we
understand better the nature of quantum gravity and its semiclassical
approximations.  Outside of that context Hadamard behavior seems
irrelevant ``operationally'', since it corresponds in the Wightman
function to the absence of a term that could only be noticed at
extremely high energies.
We hope to return to this question in another place, and to show in
particular how, by ``smoothing the boundary'', one can tweak the
ultraviolet behavior of the SJ state so that it becomes Hadamard.


\section{The massless scalar field in two-dimensional flat spacetimes}

As background for our investigation of the massless scalar field in a two-dimensional causal
diamond, we review the massless scalar field in two-dimensional Minkowski and Rindler spacetimes. The metric on Minkowski spacetime in Cartesian coordinates $(t, x)$ is given by
\be
    ds_M^2={-dt}^{2}+{dx}^{2}. \label{mink}
\ee
Since the spacetime is globally hyperbolic
and static with timelike Killing vector $\kappa_M=\partial_t$,
we can separate the
solutions to the Klein-Gordon equation into positive and negative
frequency with respect to  $\kappa_M$.  The field equation is
\be
  \Box_{M} \phi=-\partial_t^2\phi+ \partial_x^2\phi= 0\label{KG_mink}
\ee
and the normalised positive frequency modes can be taken as
\be
  {u }^M_{\small{k} } (t,x)= \frac{ 1}{ \sqrt[ ]{ 4 \pi  {\omega}_{\small{k} } }  }{e}^{-i \omega_{\small{k} }  t+ikx},\label{modes}
\ee
where $\omega_{k}=|k|$.

If we try to define a vacuum state, $| 0_M \rangle$ in
the usual way as the state annihilated by the
operator coefficients of the positive frequency modes in the
expansion of the field operator $\hat\phi(t,x)$,
then it is well-known that
we encounter
an infrared divergence ~\cite{Coleman:1973ci, Abdalla:1991vua, Faber:2002ve}:
\be
\bal
  {W}_{M}(t,x;t',x')
  &:=\langle 0_{M}| \hat\phi  (t,x  )\hat\phi  ( t',x'    )  |0_{M}  \rangle\\
  &\;=\frac{ 1}{4\pi } \int_{-\infty }^{\infty }  \frac{dk }{ |k|} {e}^{-i |k|(t-t') + ik(x-x') },
\eal
\ee
which
is logarithmically divergent at $k=0$.

Following \cite{Abdalla:1991vua},
we can remove the divergence by
introducing an infrared momentum cutoff $\lambda$
\be
 \begin{split}
 {}&\frac{ 1}{4\pi } \int_{-\infty }^{\infty }  \frac{dk }{ |k|} {e}^{-i |k|\Delta t + ik\Delta x }\theta(|k|-\lambda)\\
 &=\frac{ 1}{4\pi }  \lim_{  \epsilon  \rightarrow 0^{+}}
  \int_{ \lambda }^{  \infty }  \frac{ dk}{ k  }
  \left[ {e}^{-ik(\Delta t +\Delta x-i \epsilon) }+{e}^{-ik(\Delta t - \Delta x - i\epsilon)}\right]\\
&=-\frac{ 1}{4\pi }  \lim_{  \epsilon  \rightarrow 0^{+}}\left[\vphantom{a^2}\Log\left[i(\Delta t +\Delta x-i\epsilon)\mu\right]+\Log\left[i(\Delta t -\Delta x -i\epsilon)\mu\right]\right]+\mathcal O(\lambda\Delta)\\
&=-\frac{ 1}{2\pi } \Log\mu|d|-\frac{i }{4 }\text{sgn}(\Delta t )\theta(\Delta t^{2}-\Delta x^{2})+\mathcal O(\lambda\Delta), \label{Wight_Mink_cutoff}
\end{split}
\ee
where $\mu= \lambda e^\gamma$,
$\gamma$ is the Euler-Mascheroni constant,
and
$\Delta t=t-t'$,
$\Delta x=x-x'$,
and
$d=\sqrt{-\Delta t^{2}+\Delta x^{2}}$.
The logarithm here is given a branch cut on the negative real axis and
$\Log$ denotes its principal value.
The quantity $\Delta$ here stands
collectively for $\Delta t$ and $\Delta x$,
such that small $\lambda\Delta$
implies that
both coordinate distances $\Delta t$ and $\Delta x$ are small compared to $\lambda^{-1}$.
With non-zero $\lambda$, the theory has a preferred frame.  However, if
we drop the $\mathcal{O}(\lambda)$ term in \eqref{Wight_Mink_cutoff}, we
obtain a one-parameter family of two-point functions
that depend on $\lambda$ but are fully frame-independent:
\be
  {W}_{M, \lambda}(t,x;t',x')
  := -\frac{ 1}{2\pi } \Log\mu|d|-\frac{i }{4 }\text{sgn}(\Delta t )\theta(\Delta t^{2}-\Delta x^{2})
\label{Wight_Mink}
\ee
Unfortunately, \eqref{Wight_Mink} cannot itself serve as a physical Wightman function, because
it fails to be positive as a quadratic form (this being condition (ii) above).
Nevertheless we will see that the form \eqref{Wight_Mink} will emerge in a natural manner as a
certain limit of the two-point function we will derive for the diamond.\footnote%
{Perhaps \eqref{Wight_Mink} could also be understood as defining an ``approximate state''
 valid when $\Delta t$ and $\Delta x$ are small compared to the IR scale $\lambda^{-1}$.}

It is worth noting that the theory whose fundamental field is the gradient of
$\phi$ rather than $\phi$ itself is not infrared divergent, and in fact the
vacuum expectation value
\be
\langle 0_M| \nabla_\mu{\hat\phi}  (t, x  )\hat\phi  (t', x'    )  |0_M  \rangle
= \frac{\Delta x_\mu}{2 \pi  (\Delta t^{2}-\Delta x^{2})}
\ee
already converges, except for the singularity on the lightcone. 

\begin{figure}[t]
\begin{center}
\includegraphics[width=0.4\textwidth]{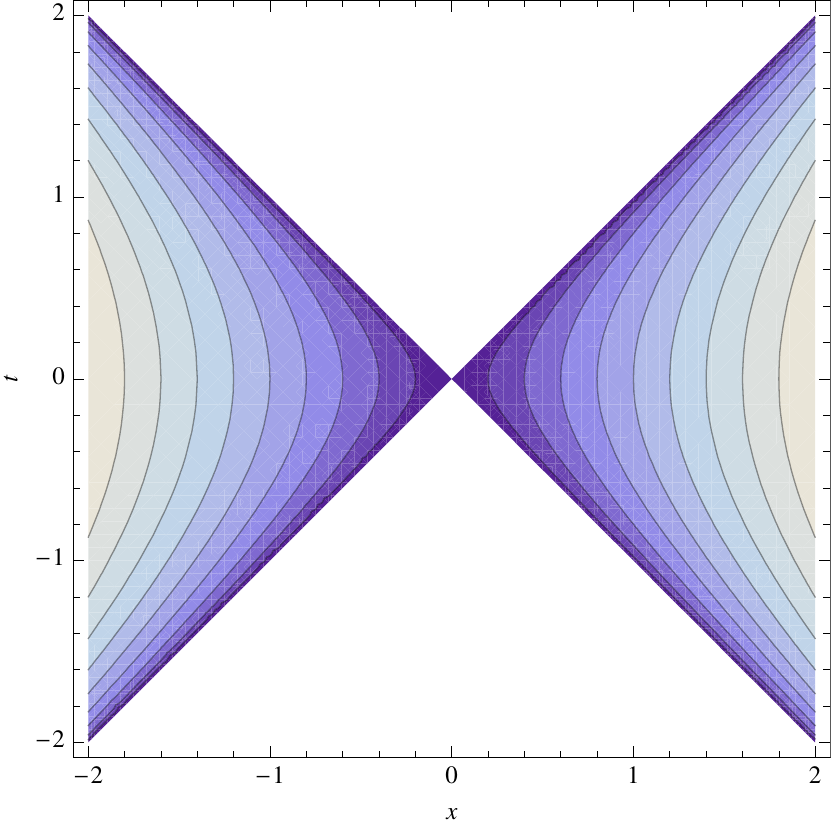}
\caption{The right and left Rindler wedges of two-dimensional Minkowski spacetime.}
\label{wedge}
\end{center}
\end{figure}

The Rindler metric~\cite{Rindler:1966zz, Rindler:2006km} arises from the Minkowski
metric via the coordinate transformations
$t=a^{-1}e^{a\xi} \sinh a\eta$ and $x=a^{-1}e^{a\xi} \cosh a\eta$,
where $a>0$ is a constant with dimensions of inverse length
and $-\infty<\xi, \eta<\infty$.
The coordinates
$\xi$ and $\eta$
only cover a submanifold of the full Minkowski space,
namely the right Rindler wedge, $x>|t|\,;$
but this submanifold is conformal to all of  Minkowski space
as one sees from the form of the line element in
$\xi$-$\eta$
coordinates:
\be
   ds_R^2=e^{2a\xi}\left(-{d\eta}^{2}+{d\xi}^{2}\right).\label{rindler}
\ee
Lines of constant $\xi$ are hyperbolae that correspond to the trajectories of observers
accelerating eternally at a constant acceleration $a e^{-a\xi}$ (figure~\ref{wedge}), and
are integral curves of the Killing vector $\kappa_R=\partial_\eta$.\\

Since Rindler spacetime is globally hyperbolic and static in its own right, the canonical
quantisation of the scalar field can be carried out in a completely self-contained
manner~\cite{Fulling:1972md}.
Thanks to the conformal invariance of the massless theory in two
dimensions, the field equation $\Box_{R} \phi= 0$ in Rindler coordinates~\eqref{rindler} is just
the usual wave equation,
whose normalised positive frequency solutions, the
Fulling-Rindler modes, are given by the plane waves
\be
  {u}^R_{\small{p}} =\frac{1}{{\sqrt{4\pi\omega_{p}}}}{e}^{-i \omega_{p}\eta+i p \xi} \ ,
  \label{16}
\ee
where $\omega_{p}=|p|$.
Trying to define a Fulling-Rindler vacuum state
$|0_R\rangle$
as the state annihilated by the operator coefficients of the positive frequency Rindler modes
in the expansion of $\hat\phi$
once again
results in an infrared divergent integral
for the two-point function $W$.
Proceeding as before,
one can introduce a
cutoff at small ``momentum'' $\lambda$ and discard a term of order $\lambda$
to obtain a
one-parameter family of two-point functions, which are the
same functions of Rindler coordinates as
the Minkowski two-point function \eqref{Wight_Mink}
is of Cartesian coordinates:
\begin{eqnarray}
 {W}_{R,\lambda}(\eta,\xi;\eta',\xi')
  & := & -\frac{1}{4\pi} \Log\mu^2|\Delta\eta^{2}-\Delta\xi^2|
        - \frac{i}{4}\text{sgn}(\Delta\eta)\theta(\Delta\eta^{2}-\Delta\xi^2)
\label{20}
\end{eqnarray}
where $\mu  = \lambda e^\gamma$.
As before,
this two-point function is symmetric (boost-invariant)
but fails to be positive and depends explicitly on the cutoff $\lambda$.
At best it
can have an approximate validity when
the coordinate-differences $\Delta\eta$ and $\Delta\xi$
are small compared to $\lambda^{-1}$.

It is well known that the Minkowski and Fulling-Rindler ``ground states''
corresponding to \eqref{Wight_Mink} and \eqref{20}
are not equal,
since the Rindler mode functions
$u^R_k$ are linear combinations of both positive and negative frequency Minkowski mode functions
$u^M_k$ and $u^{M*}_k$.
This phenomenon is by now well understood as an instance of the
Fulling-Davies-Unruh effect~\cite{Fulling:1972md,Davies:1974th,Unruh:1976db}: if the Rindler
wedge is understood as a subregion of Minkowski space,
and the field is in the usual Poincar\'e-invariant vacuum state
(say in $3+1$ dimensions, where the latter is well defined),
observers that are confined to the wedge and that accelerate
eternally at a uniform rate will feel themselves immersed in a thermal bath of particles. \\

The preceding calculations suffer from the appearance of infinite integrals and the consequent need
for infrared cutoffs.  Nevertheless, we will see that the two-point functions we have obtained in
this section can be related to the case we study next, that of a ~\emph{finite} two-dimensional
causal diamond, where the the SJ construction and the integrals it gives rise to are completely
well-defined.
In this connection we comment also that inasmuch as both the Minkowski and Rindler spacetimes
possess globally timelike Killing vectors, the formal demonstration of section \ref{sec_SJ} would
apply to show that the SJ vacua of these two spacetimes would
coincide
with the ground-states
discussed in this section -- were such states actually to exist.  (See also \cite{Siavash}.)

\section{The massless SJ two-point function in the flat causal diamond}\label{sec_diamond}

A causal diamond (or Alexandrov open set) is the intersection of the chronological future of a
point $p$ with the chronological past of a point $q\succ p$.
Because such a
causal diamond is a globally
hyperbolic manifold in its own right, the scalar field possesses therein a unique
Pauli-Jordan function $\Delta$.
In this section we will follow the SJ procedure to
derive from $\Delta$ a two-point function $W$ for the massless
scalar field in a causal diamond in two-dimensional Minkowski space.
We will then analyse the limit of $W$ for points
(i) in the centre of the diamond and
(ii) in the corner of the diamond.\\
\begin{figure}[t]
\begin{center}
\includegraphics[width=0.4\textwidth]{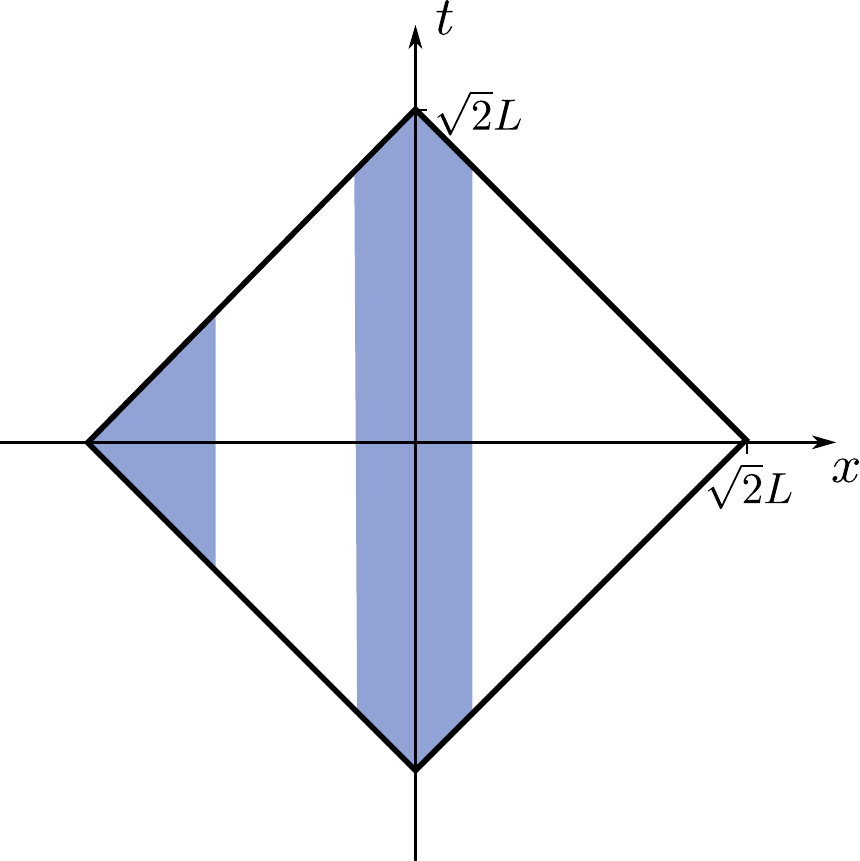}
\caption{The causal diamond, $u,v\in(-L,L)$.  The shaded portions of the diagram represent the centre $(i)$ and corner $(ii)$ regions of interest in section~\ref{sec_results}.}
\label{diamond}
\end{center}
\end{figure}

Lightcone coordinates $u=(t+x)/\sqrt{2}$ and $v=(t-x)/\sqrt{2}$ are most convenient in this
section. A causal diamond centred at the origin $u=v=0$ that corresponds to the region $u,v \in
(-L,L)$ is shown in figure~\ref{diamond} and will be denoted by $C_L$. Its spacetime volume is
$V=4L^2$.
The commutator function $i \Delta(X,X')$ can be calculated in lightcone coordinates
from the retarded and advanced propagators~\cite{bogoliubov1960introduction}:
\be
 i \Delta(u,v;u',v')= -\frac{ i}{2 } [ \theta (u-u')+\theta (v-v')-1] \ .
\label{36}
\ee
Given that $C_L$ is a bounded region of spacetime, the associated integral operator $i\Delta$
defined in~\eqref{eq:defpjfunctional} is of \emph{Hilbert-Schmidt} type, i.e.~its Hilbert-Schmidt
norm is finite:
\be
 \int_{-L }^{ L} du\int_{-L }^{ L}dv \int_{-L }^{ L}du' \int_{-L }^{ L}dv'{|i\Delta(u,v;u',v')|}^{2}=2{L}^{4}<\infty.
 \label{eq:HilbertSchmid}
 \ee
Since $i\Delta(X,X')=[i\Delta(X',X)]^*$, $i\Delta$ is also self-adjoint and so the spectral
theorem applies. The positive eigenfunctions $T^+_k$ that satisfy $i\Delta T^+_k=+\lambda_k T^+_k$,
are given by the two sets~\cite{Johnston:2010su}
\begin{align}
f_k(u,v) &:= e^{-iku} - e^{-i k v}, & &\textrm{with } k = \frac{n \pi}{L}, \; n = 1, 2, \ldots\label{eq:SJfunctions1}\\
g_k(u,v) &:= e^{-iku} + e^{-i k v} - 2 \cos(k L), & &\textrm{with } k\in\mathcal{K}, \label{eq:SJfunctions2}\end{align}
where $\mathcal{K}=\left\{k\in\mathbb{R}\,|\,\tan(kL)=2kL\textrm{ and } k>0\right\}$
and their
eigenvalues are $\lambda_k=L/k$.\footnote%
{There is a sign error in the commutator function in \cite{Johnston:2010su}
which results in the $f_k$ and $g_k$ given here being the complex conjugates of those there.}
Their $L^2$-norms are $||{f}_{\small{k}}||^{2}=8{L}^{2}$ and
$||{g}_{\small{k}}||^{2}=8{L}^{2}-16{L}^{2}{\cos}^{2}(kL)$.  It is clear that the eigenfunctions
$T^-_k$ with negative eigenvalues are given by the complex conjugates $T^-_k=[T^+_k]^*$. To verify
that the functions~(\ref{eq:SJfunctions1}-\ref{eq:SJfunctions2}) and their complex conjugates are
indeed \emph{all} the eigenfunctions with non-zero eigenvalue, we can use the
fact~\cite{Johnston:2010su,stone1979linear} that the sum over the squared eigenvalues of a
Hilbert-Schmidt operator is equal to its Hilbert-Schmidt norm.
A
short
calculation shows that
\be
\sum_{k}^{}\lambda_{\small{k}}^{2}=\sum_{n=1}^{\infty}\frac{2{L}^{4}}{{(\pi n)}^{2}}+\sum_{k\in\mathcal K}^{}\frac{{2{L}^{2}}}{k^2}=\frac{2{L}^{4}}{6}+\frac{10{L}^{4}}{6}={2{L}^{4}}
\label{46}
\ee
coincides with~\eqref{eq:HilbertSchmid}, as required (the analytic evaluation of the second sum in~\eqref{46} can be found in~\cite{Johnston:2010su, speigel1953summation}). The SJ prescription defines the two-point function ${W_{SJ,L}}(u,v;u',v')$ as the positive spectral projection of $i\Delta$:
\be
{W}_{SJ,L}(u,v;u',v')= \sum_{ n=1}^{  \infty }  \frac{ {L}^{2}}{ \pi n}  \frac{1 }{||{f}_{\small{k}}||^{2}} {f}_{\small{k}}(u,v){f}_{\small{k}}^{*}(u',v')+\sum_{k\in \mathcal K}^{  }  \frac{ {L}}{ k}  \frac{1 }{||{g}_{\small{k}}||^{2}} {g}_{\small{k}}(u,v){g}_{\small{k}}^{*}(u',v').
\label{47}
\ee \\
We denote the two sums in~\eqref{47} by $S_1$ and $S_2$, respectively.\\
The first sum
\be
S_{1}= \frac{1}{8\pi} \sum_{ n=1}^{  \infty }\frac1{n} \left[{e}^{-\frac{iun\pi}{L}}-{e}^{-\frac{ivn\pi}{L}}\right]\left[{e}^{\frac{iu'n\pi}{L}}-{e}^{\frac{iv'n\pi}{L}}\right]
\label{eq:sum1}
 \ee
can be evaluated in closed form. We recognize four Newton-Mercator series which converge to the principal branch of the complex logarithm:
\be
\bal
S_{1}= \frac{1 }{8\pi } &\left\{-\Log\left[1-{e}^{-\frac{i\pi(u-u')}{L}}\right]-\Log\left[1-{e}^{-\frac{i\pi(v-v')}{L}}\right]\right.\\
&\qquad\qquad\qquad\left.+\Log\left[1-{e}^{-\frac{i\pi(u-v')}{L}}\right]+\Log\left[1-{e}^{-\frac{i\pi(v-u')}{L}}\right]\right\}.
\label{eq:firstsum}
\eal
\ee
The second sum is
\begin{figure}[t]
\begin{center}
\includegraphics[width=0.5\textwidth]{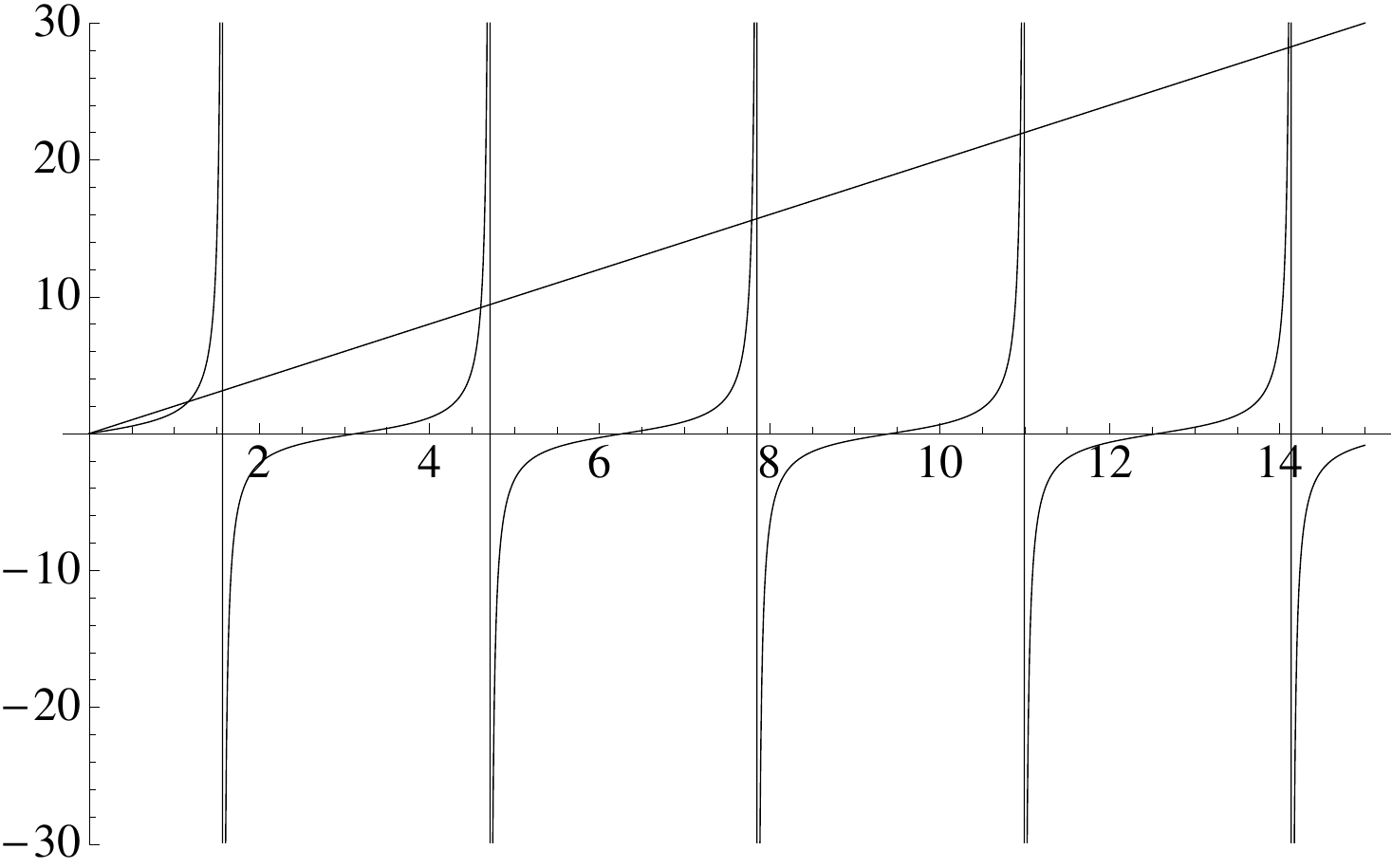}
\caption{Plot of $\tan x$ and $2x$ along with vertical lines at $\frac{2n-1}{2}\pi$ for
  $n\in\mathbb Z$.  The values of the summation variable are the positions of the intersections
  $\tan x = 2x>0$.}
\label{Transcendental plot}
\end{center}
\end{figure}
\be
  S_{2}:=
 \sum_{k\in\mathcal K}^{  } \frac{ \left[\vphantom{{e}^{iku'}}{e}^{-iku}+{e}^{-ikv}-2\cos(kL)\right]\left[{e}^{iku'}+{e}^{ikv'}-2\cos(kL)\right]}
      {kL\left[8 -16 {\cos}^{2}(kL)\right]} \ .
\label{eq:sum2}
 \ee
We have no closed form expression for this sum
but as $n\to\infty$,
the roots of the transcendental equation $\tan(x)=2x$
rapidly approach
$x_{n}=\frac{(2n-1)\pi}{2}$ with $n\in \mathbb{N}$
(see
figure~\ref{Transcendental plot}).
Therefore if we approximate the sum
by replacing $k\in \mathcal K$ with $x_n/L$,
we can expect the main error to come from a few modes of long
wavelength.  Consequently we can expect the resulting error term to be a slowly varying and small
correction to the approximated sum.
%
Let $\mathcal K_0:=\left\{\frac{2n-1}{2L}\pi\,|\,n=1,2,3\ldots\right\}$
and define the error $\epsilon(u,v;u',v')$ by
\be
\bal
S_{2}=& \sum_{k\in\mathcal K_0}^{  } \frac{ \left[\vphantom{{e}^{iku'}}{e}^{-iku}+{e}^{-ikv}-2\cos(kL)\right]\left[{e}^{iku'}+{e}^{ikv'}-2\cos(kL)\right]}{kL\left[8 -16 {\cos}^{2}(kL)\right]} + \epsilon(u,v;u',v')
\\
  = &
  \frac1{4\pi}\sum_{ n=1}^{\infty} \frac{1}{2n-1}\left[\vphantom{{e}^{\frac{iv'(2n-1)\pi}{2L}}}{e}^{-\frac{iu(2n-1)\pi}{2L}}+{e}^{-\frac{iv(2n-1)\pi}{2L}}\right]\left[{e}^{\frac{iu'(2n-1)\pi}{2L}}+{e}^{\frac{iv'(2n-1)\pi}{2L}}\right]+\epsilon(u,v;u',v')\,.
  \label{eq:epsdef}
\eal
\ee
The sums above converge to logarithmic terms when $L<\infty$ and are most conveniently expressed in the form:
\be
\bal
  S_{2}
  &= \frac{1}{4 \pi }\left\{\tanh^{-1}\left[e^{-\frac{i \pi  (u-u')}{2 L}}\right]+\tanh^{-1}\left[e^{-\frac{i \pi  (v-v')}{2 L}}\right]\right.\\
  &\qquad\qquad\;\;\;\qquad\left.+\tanh^{-1}\left[e^{-\frac{i \pi  (u-v')}{2 L}}\right]+\tanh^{-1}\left[e^{-\frac{i \pi  (v-u')}{2 L}}\right]\right\}
  +\epsilon(u,v;u',v')
  \label{eq:secondsum}
\eal
\ee

The two-point function of the SJ ground state in the causal diamond is given by the sum,
$W_{SJ,L}=S_1+S_2$.
Using
  $\tanh^{-1}(x)=\frac12\Log(1+x)-\frac12\Log(1-x)$ and the fact that
  $\Log(1-e^x)-\Log(1+e^{\frac{x}{2}})=\Log(1-e^\frac{x}{2})$,
we can combine the two sums to yield
\be
\bal
  W_{SJ,L}
  &=\frac{1}{4\pi}\left\{-\Log\left[1-{e}^{-\frac{i\pi(u-u')}{2L}}\right]-\Log\left[1-{e}^{-\frac{i\pi(v-v')}{2L}}\right]\right.\\
  &\qquad\qquad\qquad\left.
  +\Log\left[1-{e}^{-\frac{i\pi(u-v')}{2L}}\right]+\Log\left[1-{e}^{-\frac{i\pi(v-u')}{2L}}\right]\right\}+\epsilon(u,v;u',v')\\
  &=W_{\mathrm{box},L}+\epsilon(u,v;u',v'),
  \label{eq:SJbox}
\eal
\ee
where $W_{\mathrm{box},L}$ is the exact continuum two-point function of the {\it ground state} of
a massless scalar field in a box with reflecting boundaries at $x=\pm\sqrt2L$.  We shall now
investigate the form of the SJ ground state in the limits (i) and (ii) mentioned above.

\subsection{The SJ state in the centre and corner}\label{sec_results}
The limits we are concerned with require that two spacetime points
be
separated by a small
geodesic distance compared to the diamond scale $L$
and
that they be
confined to (i) the centre of the
diamond or (ii) a region near the left/right corner (we choose the left corner without loss of
generality). In these particular limits, we want the arguments in the exponentials of the sums
above to have a small magnitude so that we can Taylor expand the functions and obtain more
illuminating forms of the two-point function. Keeping in mind that in $u,v$-coordinates, the
centre of the diamond lies at $(0,0)$, the limit that corresponds to a pair of points in the centre
$(i)$ can be defined as
\be
\bal
\qquad |u-u'|&\ll L&, \qquad |v-v'|&\ll L, \qquad &|u-v'|\ll L&, \qquad &|v-u'|\ll L&.
\end{aligned}
\label{eq:restrict}
\ee
The limit that corresponds to the corner can be obtained in a similar manner by first translating the coordinate system such that the left corner of the diamond lies at the origin $(0,0)$ of the new coordinates. This corresponds to the (passive) coordinate transformation $x\rightarrow x-\sqrt{2} L$, or $u \rightarrow u- L$ and $v\rightarrow v+ L$. By ``in the corner'' we then mean the limit $(ii)$ where we first perform this translation and then apply the restriction~\eqref{eq:restrict} to the translated coordinates.\\

The inequalities
\eqref{eq:restrict} constrain the pairs of spacetime points to be separated by
a small geodesic distance $|d|=(2|u-u'||v-v'|)^{\frac12}\ll L$ and furthermore
imply
that $|x|,|x'|\ll L$.
This means that the points are
confined
to a narrow vertical strip centred
on $(i)$ the centre of the diamond
or $(ii)$ the left corner of the diamond, 
 as illustrated in figure~\ref{diamond}.
Let the width of the strip be $D\ll L$.

\subsubsection{The centre}
We first analyse the sum in the centre by expanding to lowest order in $\delta/L$, where $\delta$
collectively denotes the coordinate differences $u-u',v-v',u-v',v-u'$. Using
$\Log(1-e^x)=\Log(x)+\mathcal O(x)$ we identify the leading term in $S_1$ as
\be
S_1=\frac1{8\pi}\left[-\Log|u-u^\prime||v-v^\prime|+\Log|u-v^\prime||v-u^\prime|+C_1\frac{i\pi}{2}\right]+\mathcal O\left(\frac{\delta}{L}\right)
\ee
where $C_1=\sgn(u-u')+\sgn(v-v')-\sgn(u-v')-\sgn(v-u')$. Similarly, in $S_2$, we expand $\tanh^{-1}(e^x)=-\frac12\Log\frac{x}{2} +\mathcal O(x)$ to obtain
\be
  S_2=\frac1{8\pi}\left[-\Log|u-u^\prime||v-v^\prime|-\Log|u-v^\prime||v-u^\prime|-4\Log\frac{\pi}{4L}+C_2\frac{i\pi}{2}\right]
      +\epsilon+\mathcal O\left(\frac{\delta}{L}\right),
\ee
where $C_2=\sgn(u-u')+\sgn(v-v')+\sgn(u-v')+\sgn(v-u')$.

Now we deal with the correction, $\epsilon$.
Over a small region, $\epsilon$
should not vary much.
To investigate this, we now further
restrict the arguments of $W$ to lie
in a small square centred on the origin, of linear dimension $D$, the width of the strip.
After an
analysis and numerical investigation
given
in the appendix,
we find that
$\epsilon$ is indeed approximately constant over the small diamond and tends to a value
$\epsilon_{\mathrm{centre}}\approx -0.063$ as $L$ tends to infinity.

The terms with arguments $u-v'$ and $v-u'$ cancel in $S_1+S_2$ so we obtain:
\be
  {W}_{\mathrm{centre}}(u,v;u',v') = -\frac{1}{4 \pi}\Log|  \Delta u \Delta v|-\frac{i} {4}\text{sgn}(\Delta u+\Delta v)
   \theta(\Delta u\Delta v)-\frac1{2\pi}\Log\frac{\pi}{4L}+\epsilon_{\mathrm{centre}}+\mathcal O\left(\frac{\delta}{L}\right),
\label{eq:SJtpcentrecomplete}
\ee
as $L$ gets large.
Recall now  that $|\Delta u\Delta v|=\frac12 |d|^2$.
It is then evident that \eqref{eq:SJtpcentrecomplete}
matches the ``cut off'' Minkowski two-point function $W_{M, \lambda}$
\eqref{Wight_Mink} with a particular value of the cutoff $\lambda$
given by
\be
 \lambda=\frac{\pi}{4\sqrt2}\exp(-\gamma-2\pi\epsilon_{\mathrm{centre}})L^{-1} \approx 0.46\times L^{-1} \ ,
 \label{eq:lambda}
\ee
where $\gamma$ is the Euler-Mascheroni constant.\\

As one would expect,
$\lambda^{-1}\sim L$ for large $L$,
leading to a logarithmic factor of the form
$\Log| L^2 / \Delta u \Delta v |$.\\

Whilst strictly speaking we cannot take the $L\rightarrow \infty$ limit of such an expression,
it seems fair to say that the SJ state takes on the character of a Minkowski vacuum
in the centre of a large diamond. 
(One might wonder how this is possible given that the regularized Minkowski vacuum violates positivity but the SJ state does not. It is possible that the discrepancy is to be found in the last term of \eqref{eq:SJtpcentrecomplete}, and in any case, as already mentioned, we cannot really take the limit $L\rightarrow \infty$).
%
Notice finally that, for any $L$, 
the imaginary part of the two-point function does satisfy the requirement
$\Im(W)=\Delta/2$,
as had to be the case.

\subsubsection{The corner}
For spacetime points close to the edges of the diamond, the boundaries of the diamond appear like
the causal horizons of a Rindler wedge (see figure~\ref{wedge}).
To evaluate the SJ two-point
function in this limit, we perform the translation $u\rightarrow u-L$ and $v\rightarrow v+L$
described above, which shifts the corner of the diamond to the position $(0,0)$. We then Taylor
expand using~\eqref{eq:restrict} in the new coordinates.\\

The sums $S_1$ and $S_2$ can be evaluated as before, but the translation introduces $\pm$ signs
multiplying
integer multiples of $i\pi$.  A brief inspection shows that the first sum~\eqref{eq:sum1} is
unaltered by the translation, since only factors of $\exp(2\pi i)$ arise, while the second
sum~\eqref{eq:sum2}
picks
up
factors of $\exp(\pi i)$
that induce
sign changes in the terms involving $u-v'$ and $v-u'$.
The second sum now evaluates to
\be
  S_2=\frac1{8\pi}\left[-\Log|u-u^\prime||v-v^\prime|+\Log|u-v^\prime||v-u^\prime|+C_2\frac{i\pi}{2}\right]
  +\epsilon+\mathcal O\left(\frac{\delta}{L}\right),
\ee
where $C_2=\sgn(u-u')+\sgn(v-v')-\sgn(u-v')-\sgn(v-u')=C_1$.
The correction
term $\epsilon$ can again be analysed numerically --- see appendix A --- and the result is
that it varies very little over the small corner region for fixed $L$ and tends to
zero as $L\to\infty$.
A consequence of the sign changes is that the constant terms that
depend on $L$ cancel in $S_2$, whence there is no longer any obstruction
sending
$L\rightarrow\infty$.
Taking this limit, we obtain the two-point function
\be
\bal
\lim_{L\rightarrow \infty}{W}_{\mathrm{corner}}(u,v;u',v')=
 &-\frac1{4\pi}\Log\left|\frac{(u-u')(v-v')}{(u-v') (v-u')}\right|\\
 &-\frac{i }{4 }\text{sgn}(\Delta u+\Delta v)[\theta(\Delta u\Delta v)-\theta((v'-u) (u'-v))],
 \label{64}
 \eal
 \ee
which can be recognised as the two-point function of the scalar field in Minkowski space with a mirror
at rest at the corner $x=0$ ($x=-\sqrt2 L$ in the original coordinates)~\cite{Birrell:1982ix, Davies:1976hi}:
\be
  {W}_\text{corner}(t,x;t',x')={W}_{M,\lambda}(t,x;t',x')-{W}_{M,\lambda}(t,x;t',-x').
\label{65}
\ee
This two-point function
is scale-free
and does~\emph{not}
have the character of a
canonical vacuum for a Rindler wedge.\\

What conclusions can we draw from this?
Previously, we argued heuristically that the SJ state in the Rindler
wedge should be the Fulling-Rindler vacuum (to the extent that either is defined at
all in the presence of the IR divergences).
Now we have seen that a well-controlled limiting procedure gives a
different result.  As $L$ gets large, the spacetime geometry approaches
that of the Rindler wedge, as far as points that remain close to one
corner of the diamond are concerned, but the SJ state approaches the ground
state of a scalar field with reflecting boundary conditions at the
corner.

In fact, this ``mirror behaviour'' is already visible at the level of
the SJ modes themselves.  The first set of modes
$f_k$~\eqref{eq:SJfunctions2} vanish on the two vertical lines at the
spatial positions of the corners $f_k(x=\pm\sqrt2 L, t)=0$ (recall that
the corners are at $x=\pm\sqrt2 L$ in the coordinate system before
translation), while the second set $g_k$ also vanish on these vertical
lines in the approximation $\mathcal K\rightarrow \mathcal K_0$.  The SJ
conditions thus satisfy approximately the boundary conditions for two
static mirrors, one at each corner of the diamond.
How would such a two-mirror state appear near to the left corner?  As
the size of the diamond tended to infinity, one might expect the field
in the left corner to become unaware of the right mirror, and this is
consistent with our calculation above.  On the other hand there remains
the puzzle of where the left-hand mirror comes from in the limit.  Its
very existence selects a distinguished timelike direction, and since
this direction can only be covariantly defined by reference to the right
hand corner of the diamond, it seems difficult to avoid the conclusion
that the presence of the right corner retains its influence no matter
how large $L$ becomes!

It seems reasonable to attribute these counter-intuitive effects to our
having set the mass to zero.  As an aspect of its infrared pathology,
the massless field might be able to sense the boundaries of the finite
system, no matter how far away they are.  If this explanation is
correct, one would not expect to find the same mirror behavior for a
massive scalar field, since the mass should shield it from such long
range effects.  It would also be interesting to study massless and
massive fields in finite regions of Minkowski spacetime in $3+1$
dimensions.

\section{Comparison with the discrete SJ vacuum} \label{sec_cset}
In this section,
following~\cite{Johnston:2009fr,Johnston:2010su},
we will apply the SJ formalism to the massless scalar
field on a causal set that is well-approximated by the
two-dimensional flat causal diamond.
In the case of non-zero mass, the second of these references has shown
numerically that the mean of the discrete SJ two-point function
approximates well the Wightman function of the continuum Minkowski
vacuum.

The discrete version of the SJ prescription can be interpreted in two
ways. In the context of quantum gravity, causal sets are considered to
be fundamental --- more fundamental than continuum spacetimes which are
just approximations to the true discrete physics \cite{Bombelli:1987aa}.
On this view, the discrete SJ proposal is a starting point for building
a theory of quantum fields on the physical discrete substratum of
spacetime and one can hope that it will give us clues about quantum
gravity, just as quantum field theory in curved continuum spacetime has
done. From another viewpoint, the discrete formalism can be seen as
simply a Lorentz-invariant discretisation of the continuum formalism
and can therefore be used to test or extend the results of the continuum
theory, in particular when analytic calculations are difficult.

\subsection{Causal sets and discrete propagators}
A causal set (or causet) is  a set $\mathcal C$ together with an ordering relation $\preceq$ that
satisfies the following conditions.
It is reflexive: for all $X\in \mathcal C$, $X \preceq X$.
It is antisymmetric: for all $X,Y\in \mathcal C$, $X\preceq Y\preceq X$ implies $X=Y$.
It is transitive: for all $X,Y,Z\in \mathcal C$, $X\preceq Y\preceq Z$
implies $X\preceq Z$.
And, it is locally finite: for all $X,Y\in \mathcal C$,
$|I(X,Y)|<\infty$, where $|\cdot|$ denotes cardinality and $I(X,Y)$
is the causal interval defined by $I(X,Y):=\{Z\in \mathcal C|X\preceq Z\preceq Y\}$.
For
more details on causal set theory the reader may refer to \cite{Bombelli:1987aa, Sorkin:2003bx,Henson:2006kf}.\\

To produce a causal set $\mathcal C_{\mathcal M}$ that is the discrete
underpinning of (or approximation to, depending on the point of view) a given continuous spacetime
$\mathcal M$, we
\emph{sprinkle} points into $\mathcal M$.
A sprinkling
generates a causal set from a given Lorentzian manifold by placing points at random in $\mathcal M$
via a Poisson process with ``density'' $\rho$.
This produces a causal set whose elements are the
sprinkled points and whose partial order relation is
that of the manifold's causal relation
restricted to the sprinkled points.
The expected total number of elements in the causal set will be $N=\rho V_\mathcal M$.
A causal set generated by sprinkling provides a discretisation of $\mathcal M$ which,
unlike a regular lattice, is Lorentz invariant~\cite{Bombelli:2006nm}.
In the remainder of this paper, we will set $\rho=1$ so that area will
be measured directly by number of causet elements.
Distance and area will thus be measured in {\it natural units}.

The discrete SJ two-point function for a scalar field on a finite causal
set~\cite{Johnston:2010su, Johnston:2009fr} is constructed using the same procedure
as described above:
from the retarded Green function $G_R$,
the Pauli-Jordan operator $i\Delta$ (a finite Hermitian matrix) is
constructed and the Wightman function\footnote%
{We use a lower case $w$ for the causet counterpart of the continuum
function $W$.}
$w$ is then obtained as the positive part of
$i\Delta$.
In a finite causal set, this procedure is rigorously defined since
everything is finite.
It is of course necessary that the discrete analogue of the
retarded Green function be known, which it is for massless
\cite{Sorkin:2007qi} and massive \cite{Johnston:2008za} scalar fields
on a sprinkling into two--dimensional Minkowski spacetime.
The SJ two-point function is a complex-valued function of pairs of causal set elements,
$w:\mathcal C\times \mathcal C\rightarrow\mathbb C$;
equivalently it is a complex matrix
$w^{ij}:=w(\nu_i,\nu_j)$, where $\nu_i,\nu_j\in\mathcal C$.
%
%
When $\mathcal C$ is obtained by sprinkling, each causal set element
$\nu_i$ corresponds to a point $X_i$ in the embedding continuum
spacetime. This allows us to directly compare the values of the discrete
two-point function $w^{ij}$ and those of the continuum Wightman
function $W^{ij}:=W(X_i,X_j)$.

There is numerical evidence~\cite{Johnston:2009fr}
for a massive scalar field
that
at high sprinkling density,
the  SJ two-point
function
in the causet
approximates the known
Minkowski-space Wightman function.

We will extend this study to the massless case and compare with our
results above for the continuum SJ vacuum.
We will compare the mean of the
massless discrete two-point function $w$ with its continuum
counterparts, using two separate methods.
In the centre of the diamond,
the continuum two-point function~\eqref{eq:SJtpcentrecomplete} is
approximately a function of the geodesic distance only, in the limit of
large $L$.
We therefore plot the amplitudes $w^{ij}$ against the
proper time $|d(X_i,X_j)|$ and
ask how well they agree
with
the continuum result.
In the corner, the
continuum $W$
does not reduce,
to a function of a single variable.
In that case, we
provide a
``correlation plot''
between the discrete two-point function
and
several continuum two-point functions
(evaluated on the sprinkled
points), so that the relative goodness of fit can be assessed.

We restrict
ourselves
here to timelike
related points; the analysis for spacelike related points is
similar. Furthermore
we only need to analyze
the real parts of $W$ and $w$.
The
imaginary parts add nothing new since they are
given by the Pauli-Jordan function and
tell us nothing about the quantum state.

\begin{figure}[t]
\begin{center}
\includegraphics[width=0.4\textwidth]{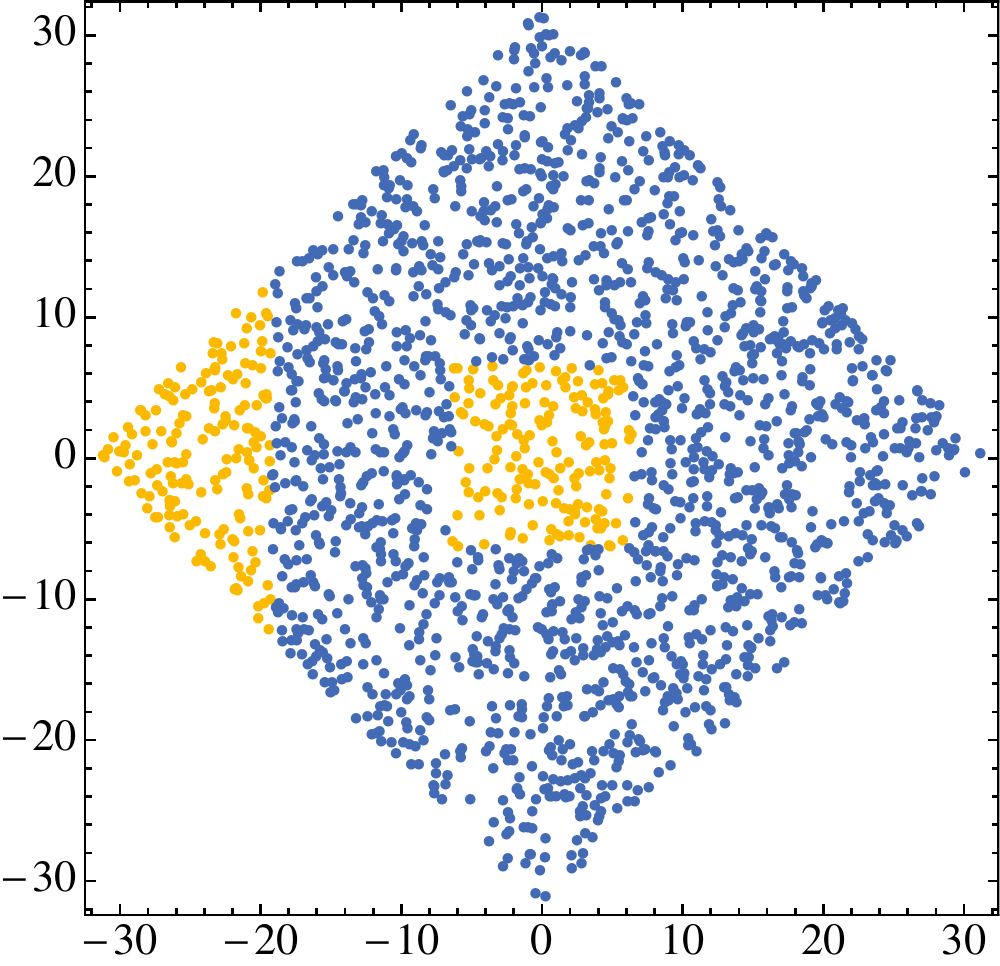}
\caption{An $N=2^{11}=2048$ element sprinkling into a diamond $C_L$ (with $L=2^{\frac92}$ in natural units). The subregions corresponding to the centre $(i)$ and the corner $(ii)$ are highlighted.}
\label{causetsprinkling}
\end{center}
\end{figure}
We
work
in a causal diamond $\mathcal M=C_L$ and evaluate the discrete propagator on an $N=2^{11}=2048$ element sprinkling
into this diamond, which implies $L=\sqrt{V}/2=\sqrt{N}/2=2^{\frac92}$ in natural units.
A typical sprinkling is shown in~figure \ref{causetsprinkling}.
Highlighted
are the two subregions in which we shall sample the discrete SJ
two-point function.  Each subregion occupies $8\%$ of the area of the
full diamond.

\subsection{The SJ vacuum in the centre}
\begin{figure}[t]
\centering
\includegraphics[width=0.8\textwidth]{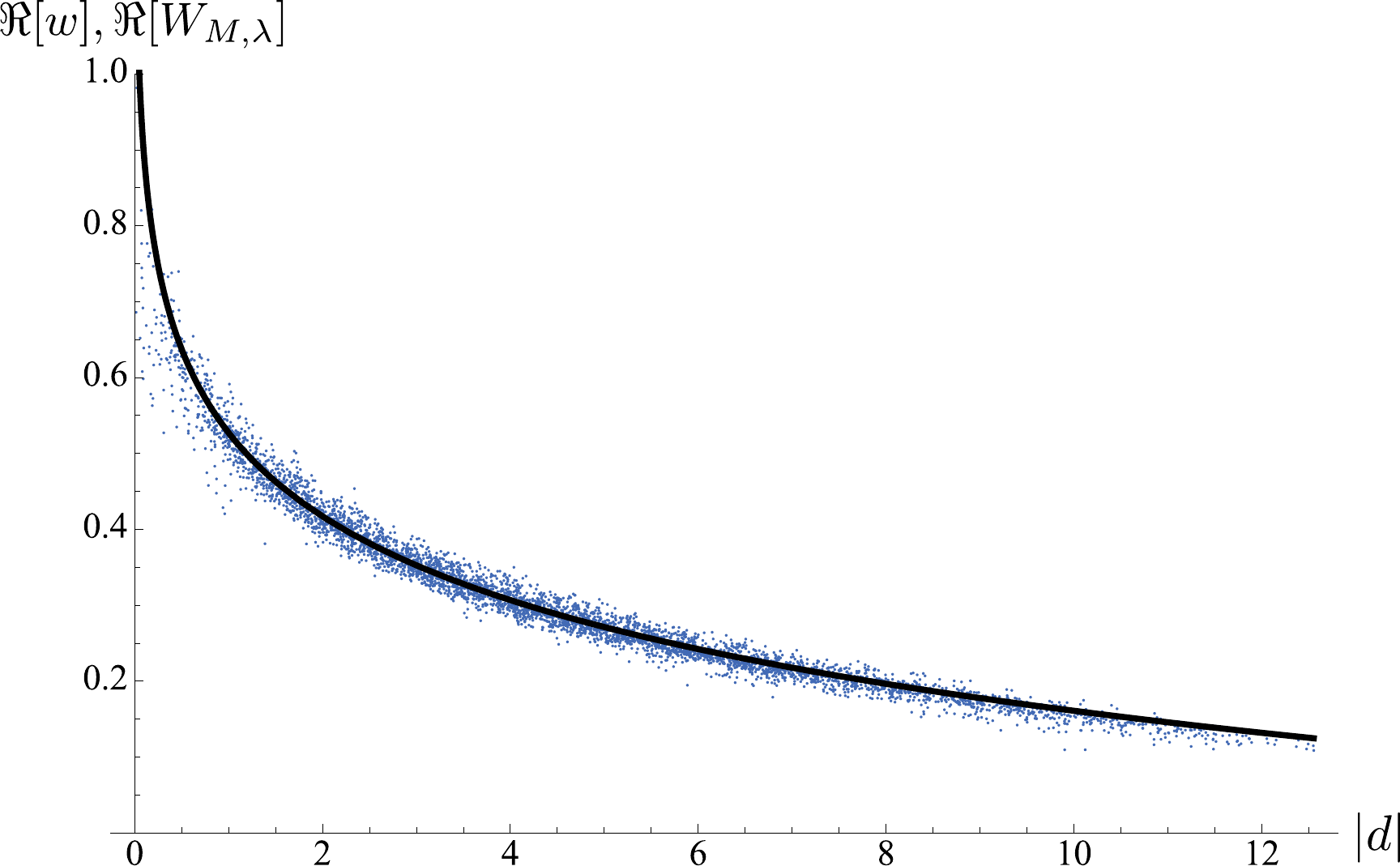}
\caption{The real parts of the continuum two-point function $W_{M, \lambda}(X,X')$ (black line) with $\lambda=0.02$ and the discrete SJ two-point function $w_{ij}$ (blue scatter) in the centre of the finite diamond $C_L$ with $L=2^\frac92$, plotted against the proper time $|d|$ for timelike separated events.}
\label{fm}
\end{figure}

The data
for
the centre is taken from a sample of 181 points in the square, among which
there were
7599 timelike related pairs.
As discussed in section 3 the two-point function of the massless
scalar field in two--dimensional Minkowski spacetime is ill-defined
owing to the infrared divergence,
but there exists a one-dimensional family
of ``approximate Wightman functions''  $W_{M,\lambda}$
parameterized by an infrared scale or ``cutoff'' $\lambda$.
It is thus natural to compare our discrete function $w$ with $W_{M,\lambda}$, where
we fix $\lambda = 0.02$ from
the
relation \eqref{eq:lambda} between $\lambda$ and $L$ when $L= 2^{\frac92}$.
Then the
real part of the continuum Wightman function we compare to is
 \be
\Re\left[W_{M,\lambda}(X,X')\right]=-\frac{1}{2\pi}\Log |d(X,X')|+0.53\,.
\label{eq:sjcentre}
 \ee
Figure~\ref{fm} displays this function
together with a scatter-plot of the discrete SJ amplitudes $\Re\left[w^{ij}\right]$
taken from
region $(i)$ in figure~\ref{causetsprinkling}.
Evidently,
the fit is good,
with a slight hint of a deviation
at larger
values of
proper time which, if real, can be attributed to
${\mathcal{O}}(\delta/L)$ corrections,
given that the centre region is still relatively large compared to
$L$.
From our previous analysis we know that
 $W_{M,\lambda}(X,X')$ approximates the continuum SJ state in the centre of the diamond,
and so
our
data also supports the conclusion that the continuum and discrete SJ
 Wightman functions approximate each other.

\subsection{The SJ vacuum in the corner and in the full diamond}

For the corner,
the type of plot we used for the centre is unsuitable because
the continuum two-point functions we want to compare with depend on more
variables than just the geodesic distance.
Instead we plot the values of $w$ directly against those of the continuum $W$ with which
we are comparing.
More specifically,
we use the coordinate values of the sprinkled points to calculate the
values of a
particular continuum function $W(X_i,X_j)=:W^{ij}$.
We then plot a point on the graph whose vertical coordinate is $W^{ij}$ and whose
horizontal coordinate is $w^{ij}$ at the corresponding pair of elements of the causet.
In this manner, we will assess the
correlation between the data sets $w^{ij}$ and $W^{ij}$ for 4 different continuum two-point
functions:
the exact continuum SJ function $W_{SJ,L}$ (before Taylor expansion),
the Minkowski function $W_{M, \lambda}$~\eqref{Wight_Mink},
the single mirror $W_{\textrm{mirror}}$~\eqref{65}, and
the Rindler function $W_{R,\lambda}$~\eqref{20}.
Both the Rindler and Minkowski  $W$-functions come with
an arbitrary parameter $\lambda$,
which
shows up on the plots as
an arbitrary vertical shift.
We set this shift such that the intercept is zero.

\begin{figure}[t]
\begin{center}
\includegraphics[width=0.24\textwidth]{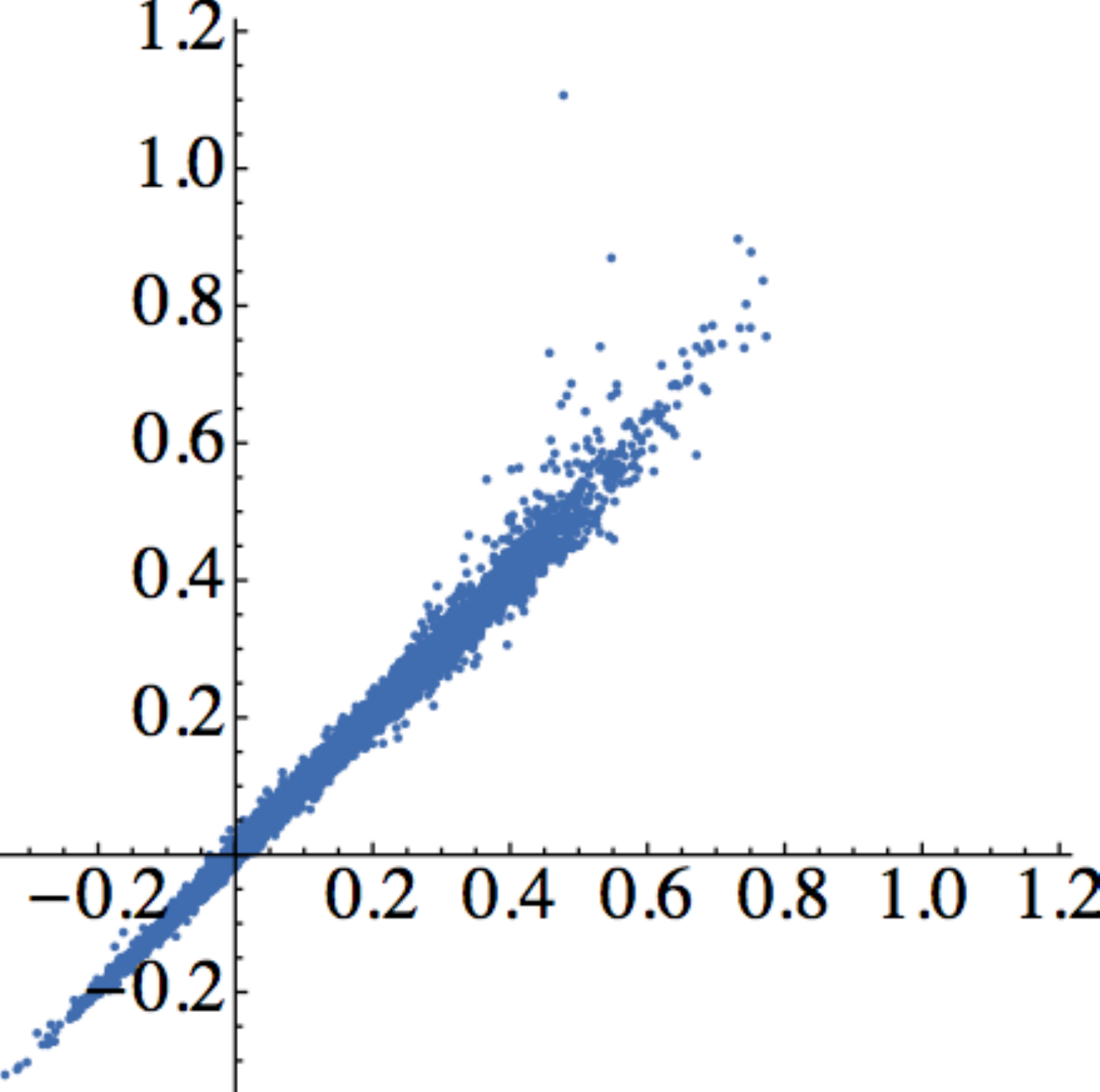}
\includegraphics[width=0.24\textwidth]{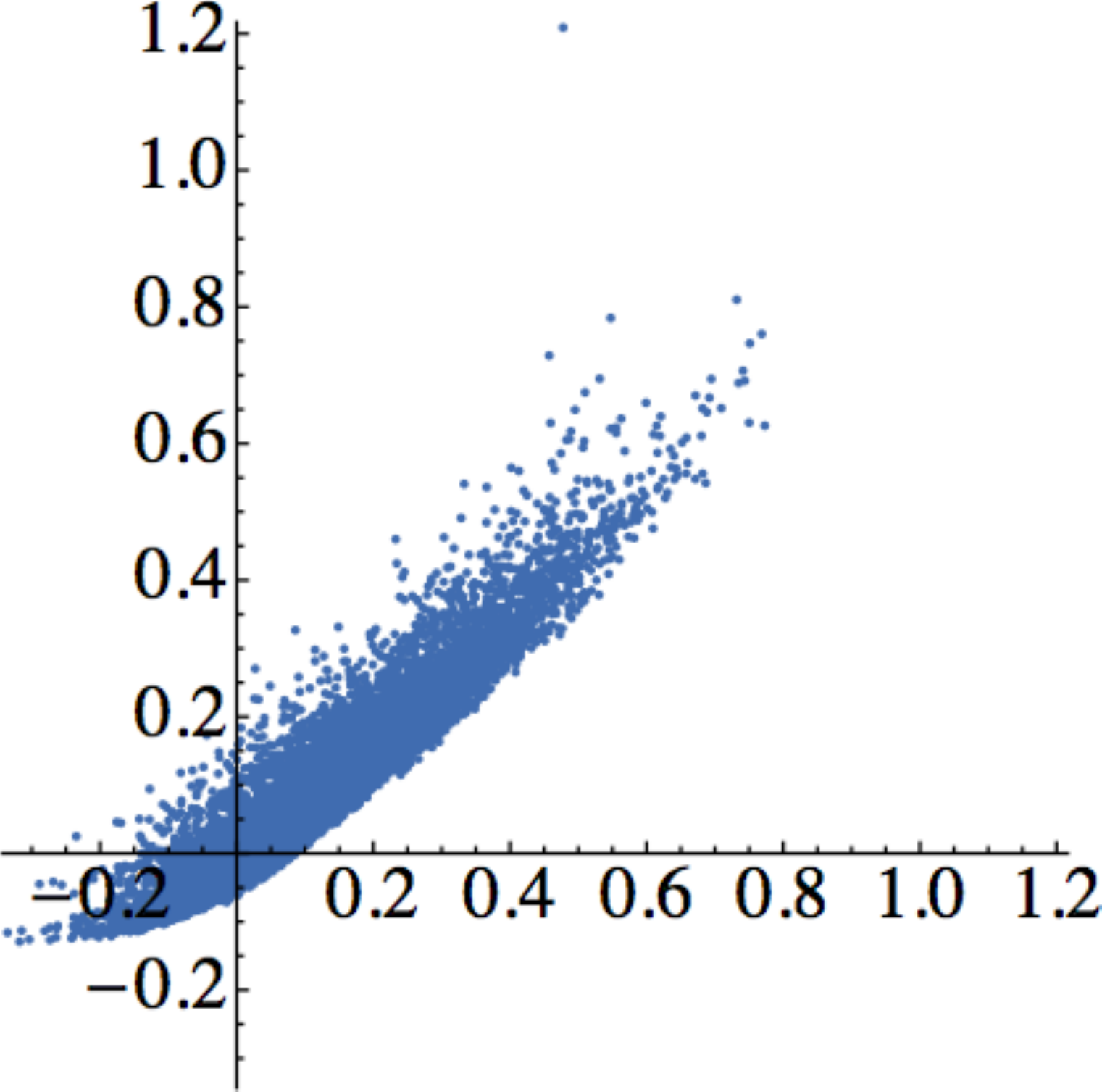}
\includegraphics[width=0.24\textwidth]{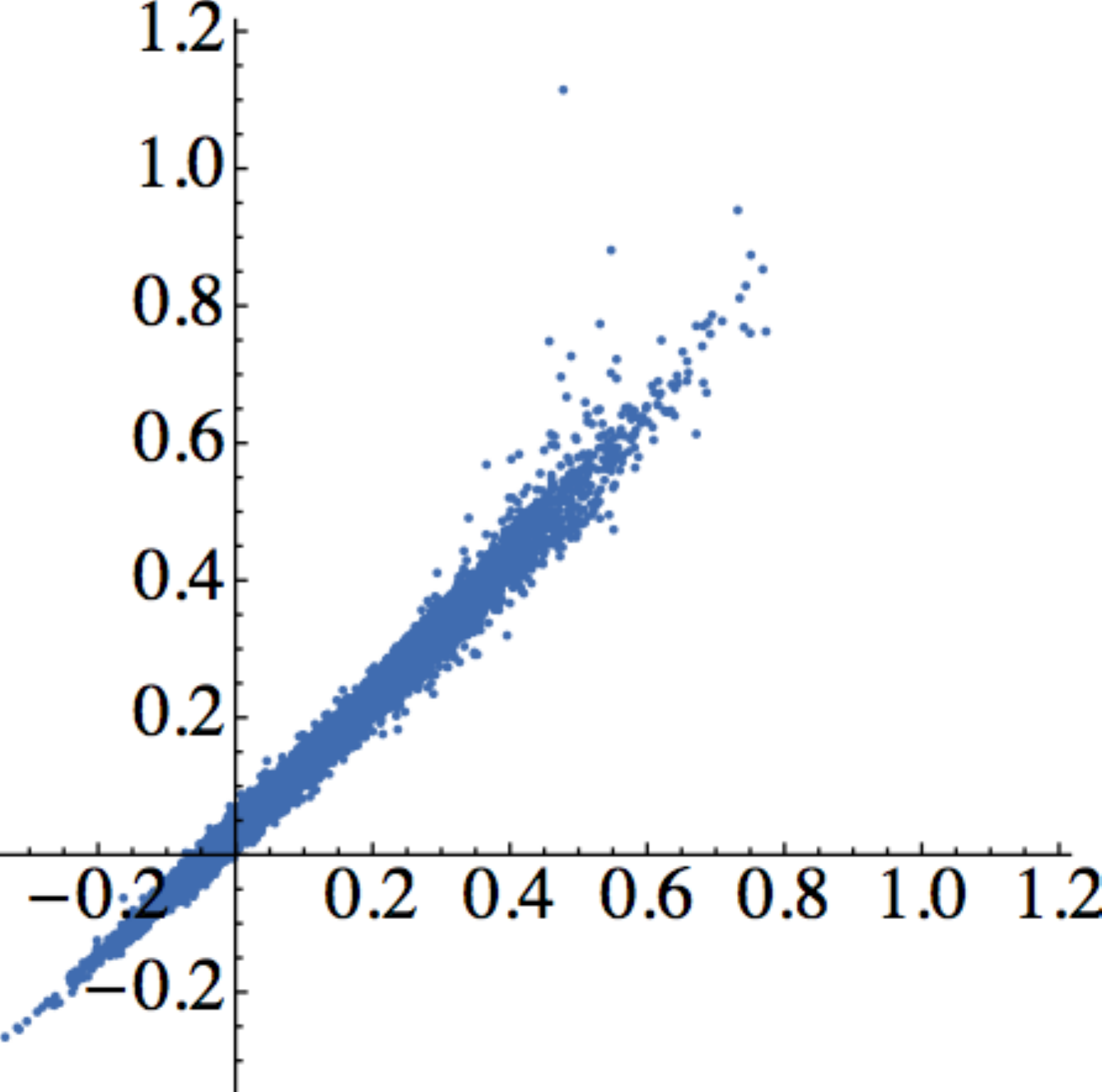}
\includegraphics[width=0.24\textwidth]{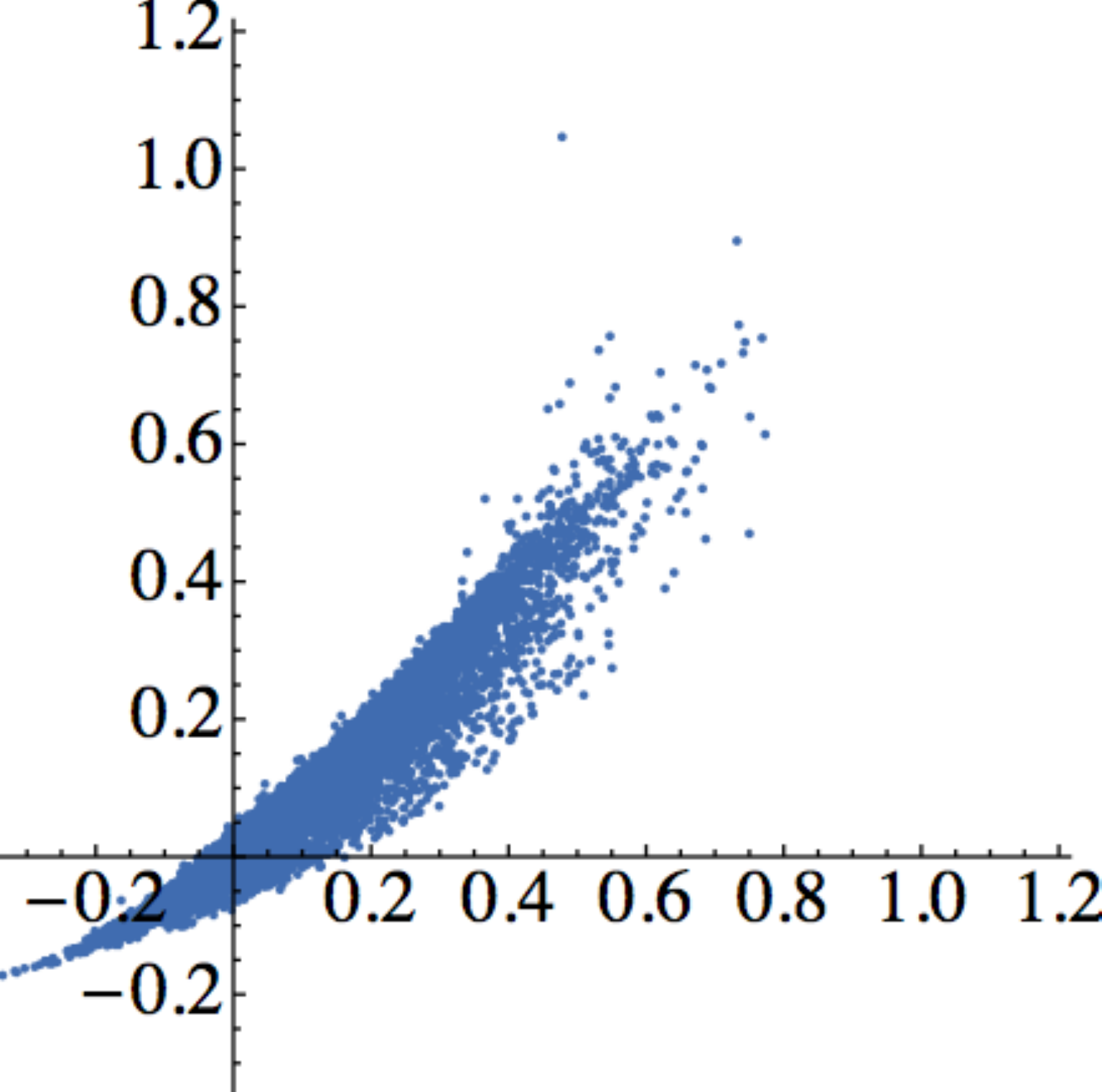}
\caption{Correlation plots for the two-point functions in the corner of the causal
diamond. The horizontal axis represents the causal set two-point function $w^{ij}$. The
vertical axes represent, from left to right: the SJ, the Minkowski, the mirror, and the
Rindler two-point functions.}
\label{fig:corrcorner}
\end{center}
\end{figure}

The corner subregion
(figure~\ref{causetsprinkling}) contains
$181$ points, which produce a sample of 11230 pairs of timelike related points.  The
correlation plots for the real parts of the discrete and continuum propagators evaluated
on this sample are shown in figure~\ref{fig:corrcorner}.
Evidently,
the
plot of the
exact SJ function exhibits a
good fit with the causal set data,
confirming again that
the discrete and
continuum formalisms agree.
As expected, the correlation with the mirror two-point
function is also high,
implying
that the ground state in the corner is indeed that of
flat space in the presence of a mirror.
%
(The slightly positive intercept in
the causal set versus mirror plot,
can plausibly be attributed to the $\epsilon$ correction and to $\mathcal{O}(\delta/L)$
effects, both of which would go away were the corner region made smaller.)
As one would expect for the corner, the Minkowski and Rindler functions exhibit
significantly worse correlations with the causet data-set.

\begin{figure}[h]
\centering
\includegraphics[width=0.192\textwidth]{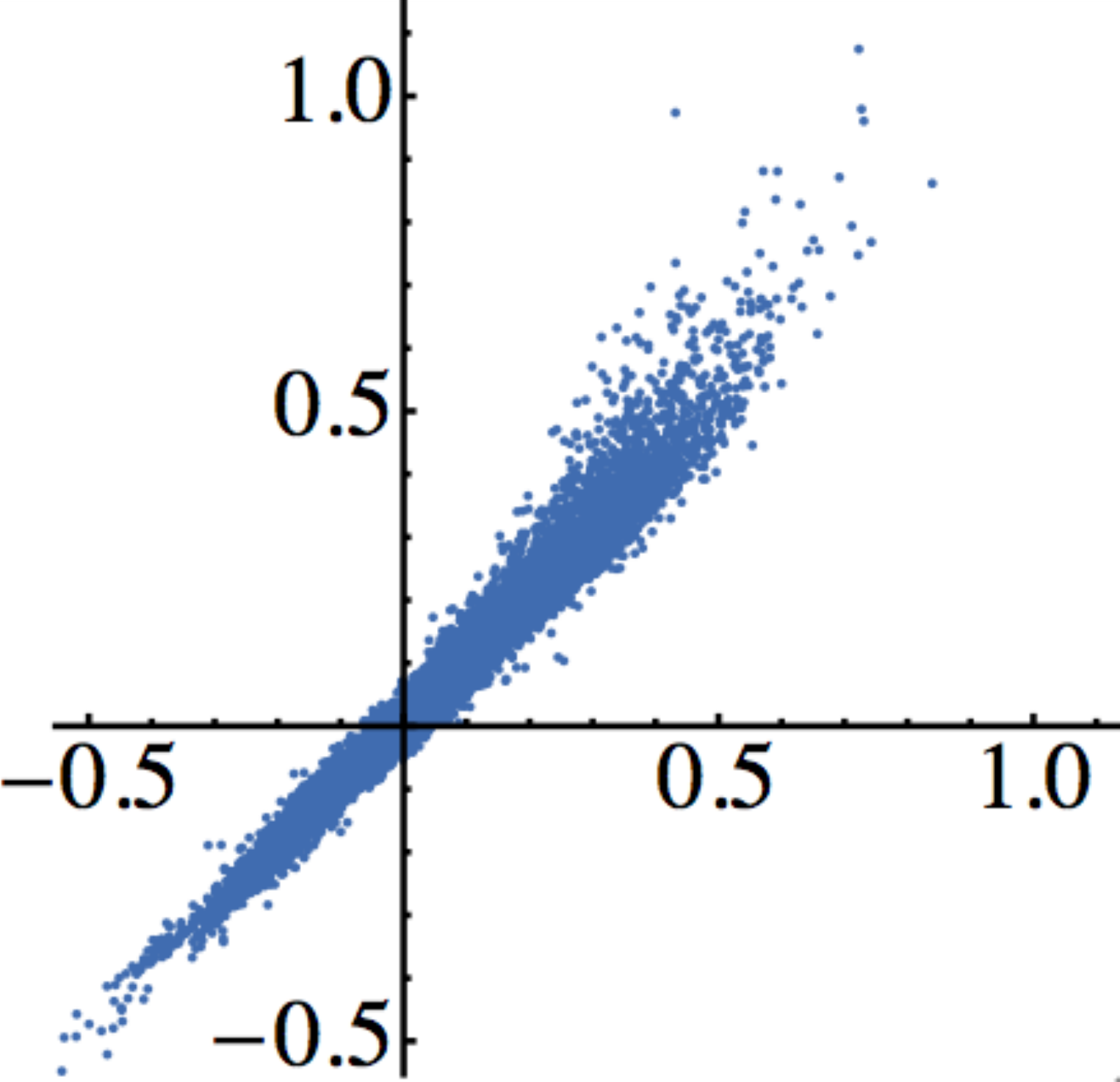}
\includegraphics[width=0.192\textwidth]{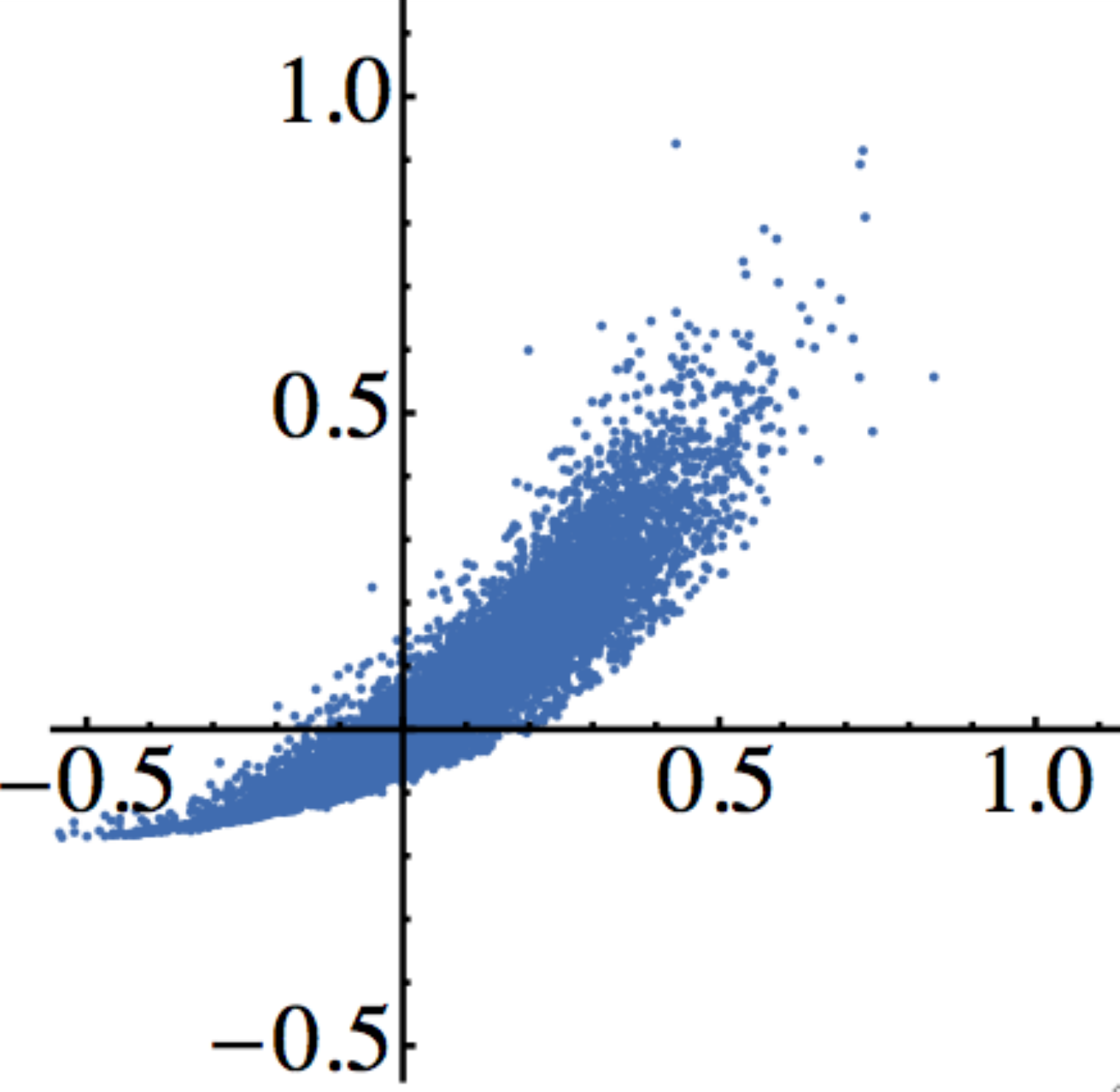}
\includegraphics[width=0.192\textwidth]{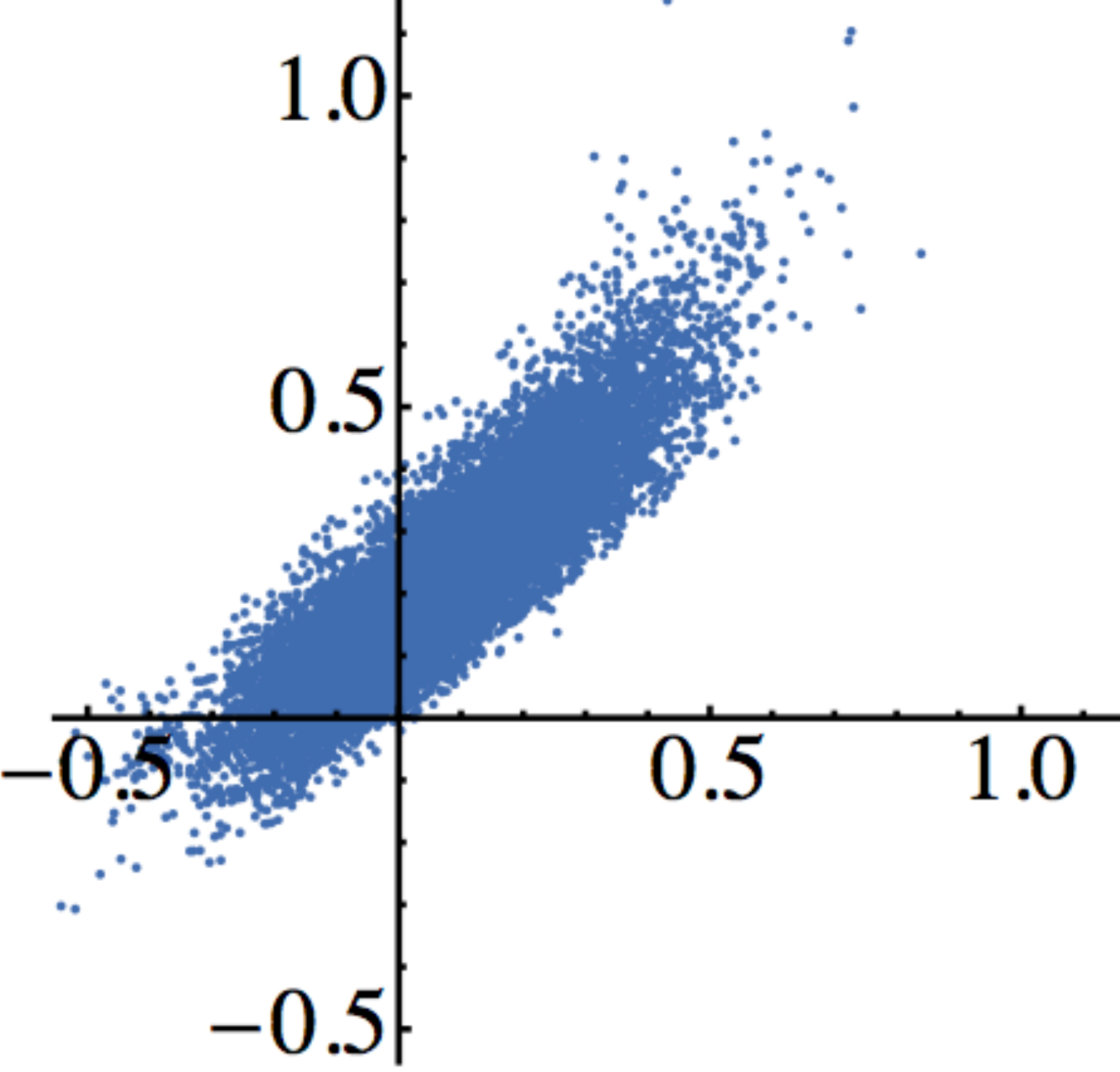}
\includegraphics[width=0.192\textwidth]{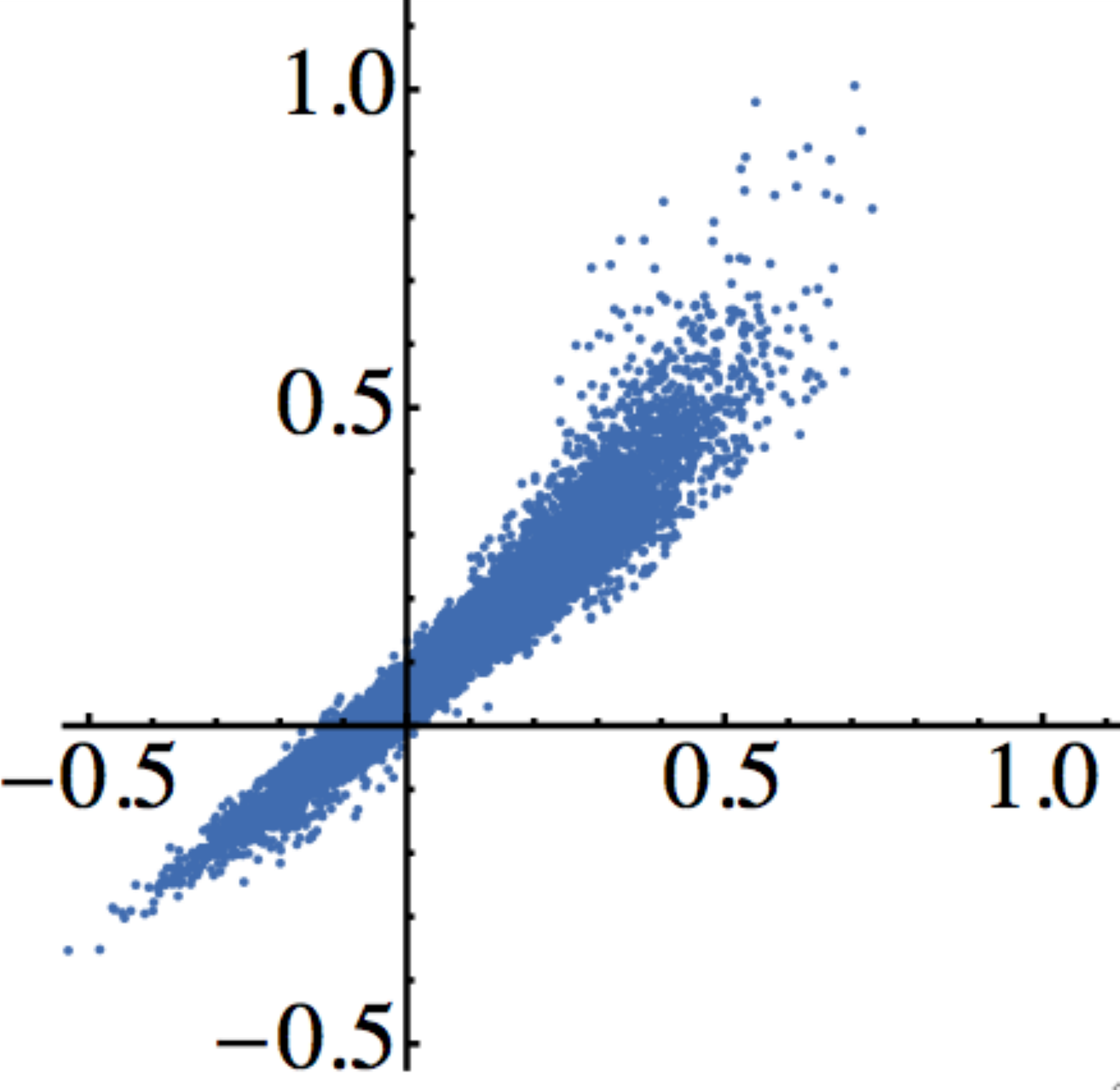}
\includegraphics[width=0.192\textwidth]{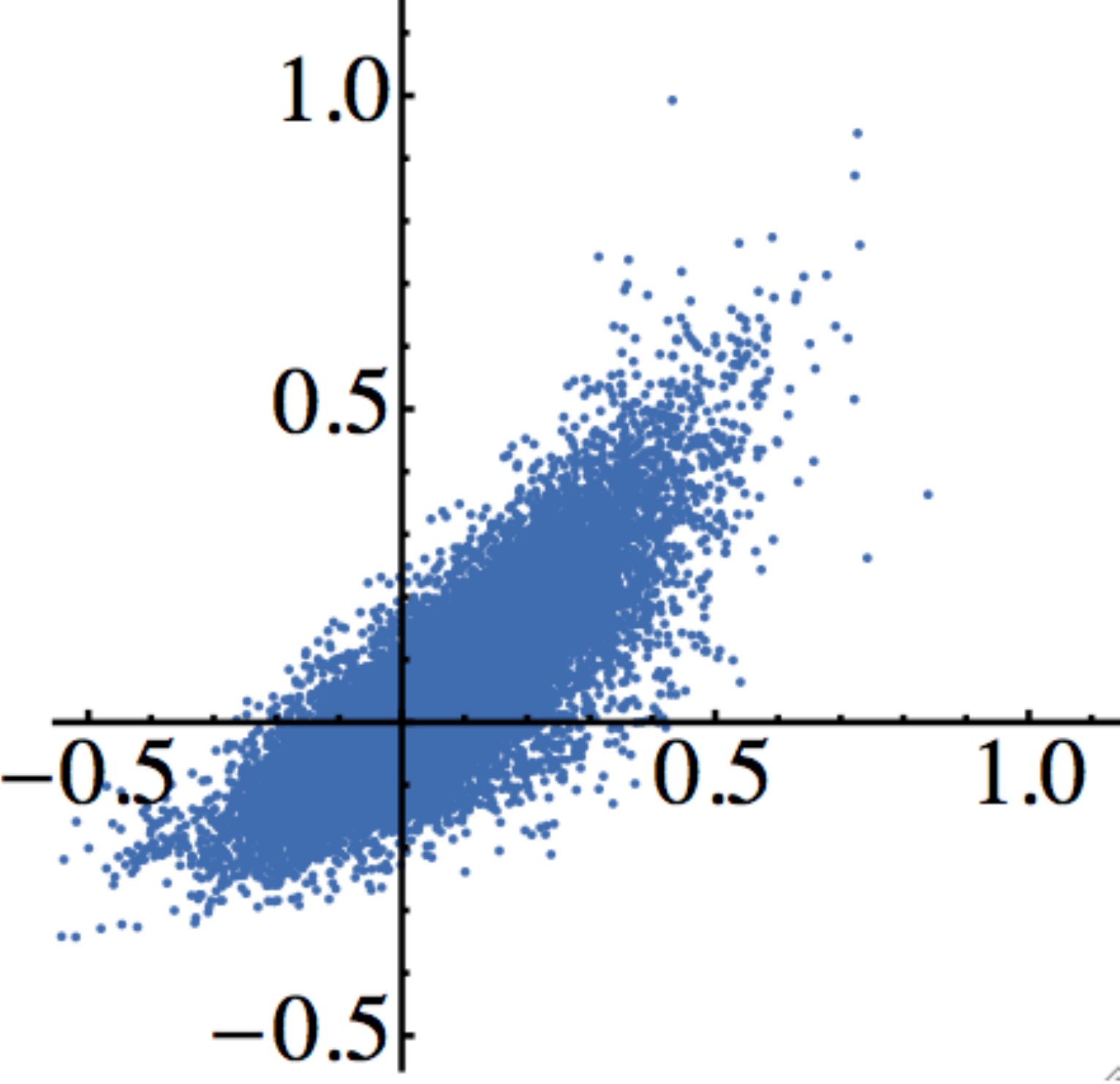}
\caption{Correlation plots for the two-point functions in the full causal diamond. The
horizontal axis represents the causal set two-point function $w^{ij}$. The vertical axes
represent, from left to right: the SJ, the Minkowski, the left mirror, the box (two
mirrors), and the Rindler two-point functions.}
\label{fig:corrfull}
\end{figure}

Turning to
the full diamond, we use a smaller overall causal set with $N=256$,
which yields a sample of $16393$ pairs of timelike related points.
To the four
comparison
functions above we add a fifth: the reflecting box (mirrors at both corners)
$W_{\textrm{box},L}$.
Of these five continuum two-point functions,
one should expect that
in addition to the SJ vacuum,
only the reflecting box ground state
would exhibit
a reasonable degree of
correlation with the causal
set data.
(As seen in equation~\eqref{eq:SJbox} above, continuum SJ and reflecting-box are
 identical up to the error-term $\epsilon(u,v;u',v')$, which, however, can vary more
 appreciably now that we do not restrict ourselves to a small subregion of the diamond.)
The correlation plots of figure~\ref{fig:corrfull} confirm this expectation,
although the match with the discrete SJ function is not a sharp as
in the case of the corner, perhaps reflecting the smaller overall sprinkling density.\\

\section{Concluding remarks}\label{sec_conclude}

We have determined the SJ vacuum of a massless scalar field in the
two-dimensional ``causal diamond'' spacetime
and compared it with suitable reference states in the full Minkowski spacetime and in the Rindler
wedge.
Both of these reference ``states'', the ``Minkowski vacuum'' and the
``Fulling-Rindler vacuum'', suffer from infrared pathologies, but to the extent that they are
meaningful we can compare them with the SJ state in the limit that the size of the diamond goes to
infinity.

When we look near the centre of the diamond, we find that the SJ Wightman function agrees with the
``Minkowski vacuum'', just as one might have expected.
However, when we look in the corner of the diamond
we do not see something having  the form of a
``Fulling-Rindler vacuum''.
Instead we recognise the flat-space vacuum
in the presence of a static mirror at that corner.
This is confirmed by the numerical results.
Thus, the  continuum calculations in section~\ref{sec_SJ}
giving the
SJ state in the
Rindler wedge do not agree with the limiting procedure of constructing the SJ state
in the finite diamond and letting the size of the diamond tend to infinity, whilst keeping
the arguments of the two-point function at fixed locations with respect to the corner.
It is important to understand the reason for this ambiguity which threatens the
proposal of the SJ state as the distinguished ground state of a region:
such a ground state, if it is defined at all, should
be unique and be independent of any (physically sensible) limiting
procedure.
Of course,
one cannot really speak of a disagreement between two functions, one of which is
ill-defined, but the fact remains that our limiting procedure yields very different
results for the centre versus the corner of the diamond.
It seems likely
that these disagreements and ambiguities
stem from the
infrared divergences of
the two-dimensional massless theory,
and if one were to work instead with a massive
field,\footnote{One might also consider working with the gradient of the field, which would be less sensitive to IR peculiarities. However, this in itself would not erase the difference between \eqref{20} and \eqref{64}, if the two point function of the gradient is the same as the gradient of the two point function.}
the SJ state for the wedge would be unique and in agreement
with both the Fulling-Rindler vacuum
and the limit of the diamond's SJ state.
Some of these questions will be investigated in future work, but preliminary numerical
results for the
massive scalar field in the sprinkled diamond indicate that
the discrete SJ function does fit
the Fulling-Rindler vacuum
better than the state with a mirror present.

If the foregoing expectations are
born out, then one conclusion would be that the SJ state for the massive field is singular on the
boundary of the diamond,
which is the behaviour one would expect for a pure state in a bounded region.
(The SJ state of a region is pure by construction.)
Indeed,
the highly entangled nature of quantum states possessing the local structure of
the Minkowski vacuum means that their
restrictions to any
spatially
bounded portion of Minkowski spacetime
will be highly mixed and far from pure.

We end by considering again the ``tug of war''
that Fulling identifies in the theory of quantum fields.  The split of spacetime into
space plus time that Fulling assumes to be inherent in the quantum aspect of the field
theory is
actually an artefact of the choice to conceive of a quantum field as if it were
the canonical quantisation of
a classical Hamiltonian system.
In a canonical approach, defining the Hamiltonian
tends to
demand the foliation of spacetime into spacelike hypersurfaces.
However, there is an alternative, long ago identified by Dirac as more fundamental because it is
essentially relativistic: the quantum analogue of the classical Lagrangian approach, namely the
path integral~\cite{Dirac:1933}.
Like the path integral,
the SJ vacuum is also relativistic in character, depending, as it does,
on the geometry of the whole spacetime region via the causal structure encoded in the
retarded Green function.
Its construction is inherently covariant,
and at no point is there a need to
introduce mode-functions defined on a hypersurface,
except as dictated by calculational convenience.
The formulation of the quantum theory along these lines
bears no resemblance to the ``canonical quantisation'' process:
one does not
solve the classical equations of motion
or identify canonically conjugate variables
or promote them to operators
satisfying canonical commutation relations.
Such a formulation seems much more compatible with a path integral approach
to quantum theory, and indeed the SJ proposal serves as the
starting point for the construction of a histories-based formulation of
quantum field theory on a causal set \cite{Sorkin:2011pn} which admits a natural
generalization to interacting scalar fields and
takes us one step further towards a quantum theory of causal sets.

\section{Acknowledgements}
We would like to thank S. Aslanbeigi and C. Fewster for helpful discussions. Research at Perimeter Institute for Theoretical Physics is supported in part by the Government of Canada through NSERC and by the Province of Ontario through MRI.
FD is supported in part by COST Action MP1006.
MB and DR thank
Perimeter Institute for hospitality during work on this paper. The work of DR was supported in
part by grant FQXi-RFP3-1018 from the Foundational Questions Institute, a
donor advised fund of the Silicon Valley Community Foundation.

\appendix
\section{Corrections to the SJ two-point function}
In
evaluating
the second sum~\eqref{eq:sum2} of the continuum SJ two-point function~\eqref{47}, we made the
approximation $\mathcal K \rightarrow \mathcal K_0$.
For a given pair of spacetime points, this
will induce an error in the two-point function, which
(see \eqref{eq:epsdef} and \eqref{47})
is given by
\be
  \epsilon(u,v;u',v')=\sum_{n=1}^{\infty}
   \left[\frac{L}{k_n}  \frac{1 }{||{g}_{\small{k_n}}||^{2}} {g}_{\small{k_n}}(u,v) {g}_{\small{k_n}}^{*}(u',v')
   - \frac{L}{k_{0,n}}  \frac{1}{||{g}_{k_{0,n}}||^{2}} {g}_{k_{0,n}}(u,v){g}_{k_{0,n}}^{*}(u',v')\right],
\label{eq:corr}
\ee
where $k_n$ and $k_{0,n}$ denote the $n^{th}$ terms in $\mathcal K$ and $\mathcal K_0$,
respectively.  The contributions to the right hand side come mostly from long wavelength (small $n$)
terms, since the approximation $\mathcal K \rightarrow \mathcal K_0$ becomes increasingly accurate
for large $n$ (see figure~\ref{Transcendental plot}). This means that we should expect $\epsilon$
to be constant over small subregions of the diamond.
We will first test this expectation numerically, restricting ourselves for simplicity to timelike
related pairs of points

To
estimate
the mean and the variation of $\epsilon(u,v;u',v')$ over different pairs of spacetime
points in a subregion associated with the centre (i) or corner (ii) of the diamond, we evaluated
$\epsilon(u,v;u',v')$ on a random sample $P$ of pairs of timelike related points within that
region.  We evaluated~\eqref{eq:corr} by truncating the sums on the right hand side at
a stage
large enough for the sum to have converged sufficiently.  We
then calculated the mean value of $\epsilon(u,v;u',v')$ and its standard deviation on the sample
$P$.  We present the results for the real part of $\epsilon(u,v;u',v')$ below, since we are mainly
interested in the real part of the Wightman function $W$.
(The imaginary part of $W$ is proportional to $\Delta$ and therefore known exactly.  Moreover, the
imaginary part of $\epsilon(u,v;u',v')$ was consistent with zero in all the regions we investigated
numerically.)
\begin{figure}[t]
\begin{center}
\includegraphics[width=0.49\textwidth]{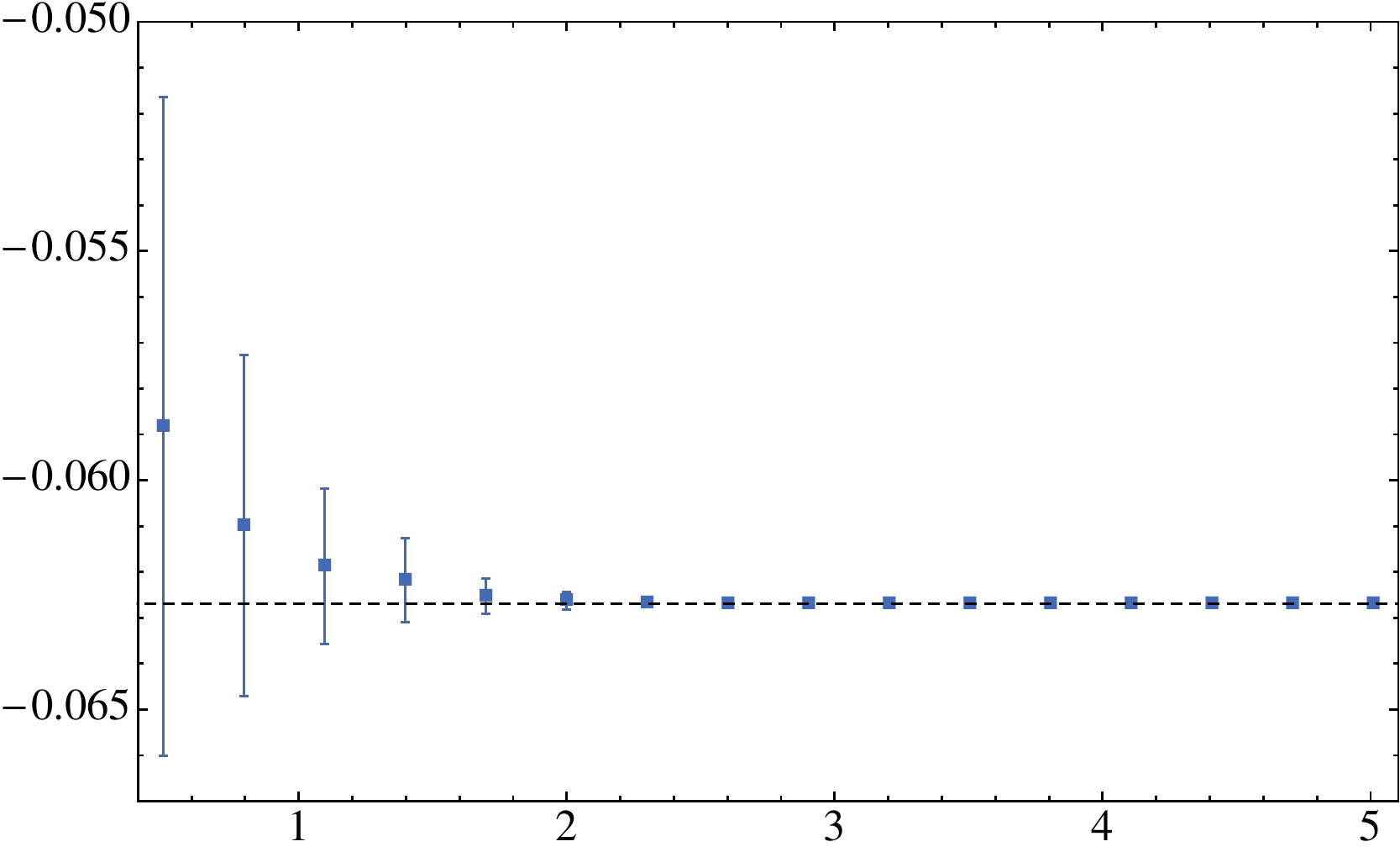}
\includegraphics[width=0.49\textwidth]{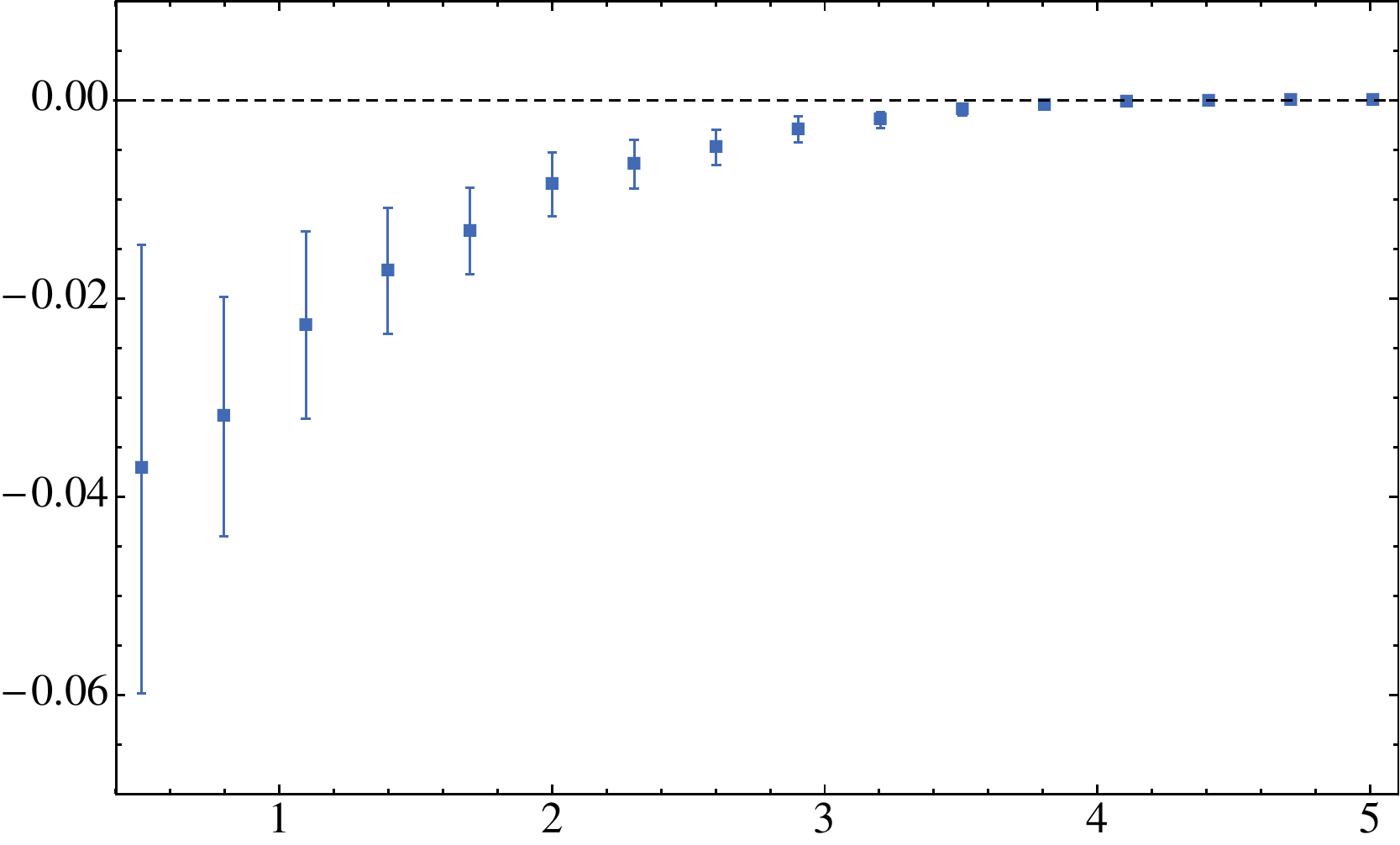}
\caption{The mean and standard deviation of $\epsilon(u,v;u',v')$ in the centre (left) and in the corner (right).  The vertical axis corresponds to $\epsilon(u,v;u',v')$ and the horizontal axis
  is $-\mathrm{log}_{10}(V_\mathrm{sub}/V)$.\label{fig:epsilon}}
\end{center}
\end{figure}

Look at subregions in the centre and corner of the form depicted in
figure~\ref{causetsprinkling}: a square in the centre, a triangle in the corner.
Fix the linear
dimension of the subregion $D\ll L$ and denote its spacetime volume by $V_{\mathrm{sub}}$.
%
Increase the size $L$ of the full diamond while keeping $D$ fixed,
thereby  decreasing
the volume ratio
$V_{\mathrm{sub}}/V$.
The mean and standard deviation of $\epsilon(u,v;u',v')$
obtained in this way
are shown in figure~\ref{fig:epsilon}
for
different values of $V_{\mathrm{sub}}/V$ ranging from $\mathcal O(10^{-1})$ to $\mathcal
O(10^{-6})$.
These
results
were obtained by truncating the
sum~\eqref{eq:corr}
at $n=50$,
which provides sufficient accuracy.
We observe that the
standard deviation in $\epsilon(u,v;u',v')$ indeed quickly becomes negligible as
$V_{\mathrm{sub}}/V$ is decreased.
The mean of $\epsilon(u,v;u',v')$ tends to a constant value in
the centre given by $\epsilon_{\mathrm{centre}}=-0.0627$ and it vanishes in the corner:
$\epsilon_{\mathrm{corner}}=0$.
Notice that these results are unchanged under a simultaneous rescaling
of $L$ and $D$: the mean and standard deviation of $\epsilon(u,v;u',v')$ depend only on the ratio
$V_{\mathrm{sub}}/V$.\\

It is worth noting that the asymptotic values for the mean of $\epsilon$ seen at large $L$ above
agree with the values of $\epsilon(u,v;u',v')$ that we obtain if
we simply evaluate the infinite sum
on a pair of coincident points in the centre of the diamond,
$(u,v)=(u',v')=(0,0)$,
or in the corner $(u,v)=(u',v')=(-L,L)$.
In the centre, the
sum~\eqref{eq:corr} reduces to
\be
\epsilon(0,0;0,0)
=\sum_{n=1}^{\infty} \left[
\frac{ \left(1-\sqrt{4x_n^2+1}\right)^2}{2x_n\left(4x_n^2-1\right)}
-\frac{1}{\pi(2n-1)}\right],
\ee
where $x_n$ is the $n^{th}$ positive solution to $\tan(x)=2x$.
This sum can be evaluated to arbitrary precision using numerical solutions for $x_n$,
and it
tends to
$\epsilon(0,0;0,0)=-0.0627$,
corresponding to the horizontal asymptote in
the centre plot of figure~\ref{fig:epsilon}.
In the corner, both terms in~\eqref{eq:corr} vanish
because the $g_k$ modes~\eqref{eq:SJfunctions2} are identically zero at $(-L,L)$ for all
$k_n\in\mathcal K$ and $k_{0,n}\in\mathcal K_0$.
It follows that $\epsilon(-L,L;-L,L)=0$,
corresponding to the horizontal asymptote in the corner plot of figure~\ref{fig:epsilon}.

\newpage
\bibliography{diamondbib}
\bibliographystyle{jhep}

\end{document}